\documentclass[aps,prd,onecolumn,superscriptaddress,nofootinbib,11pt]{revtex4}
\usepackage{epsfig,amsfonts,amsthm}
\usepackage{amsmath,amssymb}
\usepackage[utf8]{inputenc}
\usepackage[usenames,dvipsnames]{xcolor}
\newcommand{\be}{\begin{equation}}
\newcommand{\ee}{\end{equation}}
\newcommand{\bea}{\begin{eqnarray}}
\newcommand{\eea}{\end{eqnarray}}

\newcommand{\dst}{\displaystyle}
\newcommand{\fr}[2]{\frac{{\dst #1}}{{\dst #2}}}
\renewcommand{\Re}{\mathrm{Re }}
\renewcommand{\Im}{\mathrm{Im }}

\newcommand{\Tr}{\mathrm{Tr}}
\newcommand{\Z}{\mathbb{Z}}
\newcommand{\RR}{\mathbb{R}}
\providecommand{\mtrx}[1]{\begin{pmatrix} #1 \end{pmatrix}}

\newcommand{\mmmatrix}[9]{ \left(\! \begin{array}{ccc}#1 & #2 & #3\\ #4 & #5 & #6\\ #7 & #8 & #9\\ \end{array}\!\right) }

\providecommand{\id}{{\boldsymbol{1}}}

\definecolor{darkgreen}{rgb}{0.0, 0.6, 0.2}

\def\lsim{\mathrel{\rlap{\lower4pt\hbox{\hskip1pt$\sim$}}
    \raise1pt\hbox{`$<$}}}         
\def\gsim{\mathrel{\rlap{\lower4pt\hbox{\hskip1pt$\sim$}}
    \raise1pt\hbox{$>$}}}         
\newcommand{\lrpartial}{\partial^{\hspace{-6pt}\raise3pt\hbox{\small $\leftrightarrow$}}}

\DeclareMathOperator{\diag}{\mathrm{diag}}
\usepackage{hyperref}
\hypersetup{
   colorlinks=true,       
    linkcolor=Blue,          
    citecolor=Plum,        
    filecolor=magenta,      
    urlcolor=YellowOrange           
}

\begin{document}
\date{\today}
\title{
{\normalsize \hfill CFTP/19-005} \\*[7mm]
Recognizing symmetries in 3HDM in basis-independent way}

\author{I.~de~Medeiros~Varzielas}\thanks{E-mail: ivo.de@udo.edu}
\affiliation{CFTP, 
Instituto Superior T\'{e}cnico, Universidade de Lisboa,
Avenida Rovisco Pais 1, 1049 Lisboa, Portugal}
\author{I.~P.~Ivanov}\thanks{E-mail: igor.ivanov@tecnico.ulisboa.pt}
\affiliation{CFTP, 
Instituto Superior T\'{e}cnico, Universidade de Lisboa,
Avenida Rovisco Pais 1, 1049 Lisboa, Portugal}

\begin{abstract}
Higgs doublets may come in three generations. The scalar sector of the resulting three-Higgs-doublet model (3HDM) may be constrained by global symmetry groups $G$ leading to characteristic phenomenology.
There exists the full list of symmetry groups $G$ realizable in the 3HDM scalar sector
and the expressions for $G$-symmetric scalar potentials written in special bases 
where the generators of $G$ take simple form.
However recognizing the presence of a symmetry in a generic basis remains a major technical challenge,
which impedes efficient exploration of the 3HDM parameter space.
In this paper, we solve this problem using the recently proposed approach,
in which basis-independent conditions are formluated as relations among basis-covariant objects.
We develop the formalism and derive basis-independent necessary and sufficient conditions
for the 3HDM scalar sector to be invariant under each of the realizable symmetry group.
We also comment on phenomenological consequences of these results.
\end{abstract}

\maketitle

\newpage

\tableofcontents

\newpage

\section{Introduction}

\subsection{Historical context}

The scalar potential of the Standard Model (SM) minimally includes a single doublet of $SU(2)_L$ which reduces the electroweak symmetry to electromagnetism via the Brout-Englert-Higgs mechanism, see the recent review \cite{Maas:2017wzi} and references therein. The associated single physical Higgs boson has been observed \cite{Aad:2012tfa, Chatrchyan:2012ufa} and is now being extensively investigated at the LHC.
However whether the Higgs sector is indeed as minimal as postulated by the SM or if the observed 125 GeV Higgs is just the first state of a rich scalar sector is presently unknown. This question can only be answered by experiment. In anticipation of possible future hints or discoveries, theorists investigate other, non-minimal Higgs sectors and look for novel ways to experimentally probe them, see e.g. \cite{Ivanov:2017dad}.

A simple and well motivated generalisation of the SM is extending the scalar sector to include further $SU(2)_L$ doublets. This can be thought of as bringing to the scalar sector the concept of generations present in the SM fermion sector. Historically, the main motivations for going beyond the minimal scalar sector of the SM were to gain insight into the origin of $CP$ violation (CPV) and into the general flavour puzzle.

In 1973, T.~D.~Lee suggested that $CP$ can be broken spontaneously in a model with two Higgs doublets (2HDM)~\cite{Lee:1973iz, Lee:1974jb}: one starts with a lagrangian which is explicitly $CP$-invariant but observes that the vacuum expectation values (vevs) emerging after the scalar potential minimization break the symmetry.
However, one typically obtains in this case dangerously large tree-level flavor changing neutral currents (FCNCs).
Although they can be eliminated by imposing natural flavor conservation (NFC) \cite{Glashow:1976nt, Paschos:1976ay}, this extra requirement precludes any $CP$ violation, explicit or spontaneous. 
This clash was removed by S.~Weinberg in 1976 in a model with three Higgs doublets (3HDM) 
\cite{Weinberg:1976hu} with explicit CPV and later by G.~Branco in the spontaneously $CP$-violating model
\cite{Branco:1979pv,Branco:1980sz}. See also e.g. \cite{Serodio:2011hg, Varzielas:2011jr} for more possibilities to control FCNCs in $N$-Higgs Doublet Models (NHDMs).

The late 70's also witnessed a surge of activity on linking the fundamental fermion masses with the entries
of the Cabibbo-Kobayashi-Maskawa (CKM) mixing matrix. 3HDMs equipped with discrete symmetry groups
offered many intriguing opportunities.
In 3HDMs, the number of Higgs doublets matches the number of fermion generations,
which is viewed as an appealing feature of the models.
Various examples of 3HDM were constructed based on symmetry groups such as $S_3$
\cite{Derman:1977wq,Derman:1978nz,Derman:1978rx,Pakvasa:1977in}, $S_4$ \cite{Pakvasa:1978tx,Yamanaka:1981pa,Brown:1984mq}, and $\Delta(54)$ \cite{Segre:1978ji}. More details of NHDMs including further historical context can be found in \cite{Ivanov:2017dad}.

During 1990's and 2000's, exploration of 	multi-Higgs-doublet model was dominated by 2HDMs,
boosted by two Higgs doublets being required in minimal supersymmetric extensions \cite{Gunion:1989we}.
In the past decade, 3HDMs gradually re-gained interest since in many aspects 
they are capable of delivering more than 2HDMs. The attractive phenomenological
features of 3HDMs include richer scalar spectrum, CPV simultaneously 
with dark matter candidates \cite{Grzadkowski:2009bt}, geometrical CPV \cite{Branco:1983tn, deMedeirosVarzielas:2011zw},
a novel type of $CP$ symmetry, 
which is of order 4 rather than of order 2 \cite{Ivanov:2015mwl,Ivanov:2018qni,Ferreira:2017tvy}
and which is physically distinct from the usual $CP$ \cite{Haber:2018iwr},
and of course a variety of discrete symmetry groups.

Given that 3HDM scalar and Yukawa sectors can be equipped with global symmetries, 
which have a profound effect on phenomenology, a classification program was undertaken a decade ago
to list all symmetry-related situations possible in 3HDMs.
First, the list of all abelian symmetries realizable in 3HDM without leading to accidental symmetries
was obtained in \cite{Ferreira:2008zy, Ivanov:2011ae} and later extended to Yukawa sectors
in \cite{Serodio:2013gka,Ivanov:2013bka,Nishi:2014zla}.
Next, the full list of all discrete non-abelian symmetry groups realizable in the 3HDM scalar sector 
was derived in \cite{Ivanov:2012ry, Ivanov:2012fp}.
Continuous non-abelian groups were not listed;
we will include them in the present work to complete the classification.
Finally, a $G$-symmetric potential can have minima which either conserve
or (partially) break the symmetry group. The full list of all symmetry breaking patterns 
for each group $G$ was presented in \cite{Ivanov:2014doa}.
One particularly important conclusion was that, for sufficiently large discrete group $G$,
there remains some residual symmetry in {\em any} minimum.
In the light of the theorem formulated initially in \cite{Leurer:1992wg} 
and refined in \cite{Felipe:2014zka}, this incomplete breaking leads to unrealistic fermion sectors.

\subsection{The challenge of basis independent recognition: the example of $CP$ symmetry}

Models which involve several fields with equal quantum numbers
possess notorious large basis-change freedom, which can seriously impede their efficient exploration.
Two models may look completely different and in fact correspond to the same physics, merely written 
in different bases. A model can also contain a symmetry, but if its lagrangian is written in a generic basis,
the presence of this symmetry may be obscured.
In order
to detect the presence of symmetries, one must develop and apply 
symmetry recognition checks which do not rely on the choice of basis.

The traditional basis-invariant approach to NHDMs with symmetries is best illustrated by the problem
of finding necessary and sufficient conditions of explicit $CP$-conservation in the scalar sector.
In order to understand the properties of the potential 
under the action of a general $CP$ transformation \cite{Ecker:1987qp,Grimus:1995zi,Weinberg:1995mt},
one constructs $CP$-odd invariants (CPI), first identified in~\cite{Mendez:1991gp} and further developed in \cite{Lavoura:1994fv, Botella:1994cs, Branco:2005em, Davidson:2005cw, Gunion:2005ja, Varzielas:2016zjc}.
One writes the coupling coefficients of the scalar potential as tensors under the basis change group, then fully contracts these tensors to produce various basis invariant quantities, and selects those invariants
which flip sign under the action of a general $CP$ transformation. 
Although the explicit expression of the general $CP$ transformation is basis-dependent,
its action on basis invariants is the same in all bases, and therefore one gets an unambiguous identification
of CPIs. 

Although there are infinitely many CPIs, there exists a finite number of ``generating'' CPIs. 
If all of these generating CPIs are zero, then all other CPIs are also zero,
and the model is explicitly $CP$ conserving.
One just needs to identify these generating invariants, and this is where the problem becomes difficult.

In the case of 2HDM, the four generating CPIs were established in \cite{Branco:2005em, Davidson:2005cw, Gunion:2005ja} with the aid of computer algebra. They were almost immediately derived in a much more transparent way 
within the bilinear formalism, which appeared first in \cite{Nagel:2004sw}
and which was developed further and applied to $CP$-conservation in 
\cite{Ivanov:2005hg,Maniatis:2006fs, Ivanov:2006yq, Maniatis:2007vn, Ivanov:2007de}.
Very recently, the four CPIs of 2HDM were rederived in an alternative approach
based on fields rather than bilinears \cite{Trautner:2018ipq}.

Extension of these methods to 3HDM turned out very challenging.
Although the CPIs can be easily constructed \cite{Varzielas:2016zjc},
it is unclear how to find the set of generating CPIs. It was done, for example,
in simpler cases of 3HDMs with non-abelian symmetries with triplets \cite{deMedeirosVarzielas:2017glw, deMedeirosVarzielas:2017ote}, but it remains unsolved in the general 3HDM.
Whether the methods of \cite{Trautner:2018ipq} can be generalized to 3HDM and solve this problem
remains an open question and requires additional work.

Meanwhile, an alternative approach made its debut in 2006 \cite{Nishi:2006tg}
and was recently exploited fully in \cite{Ivanov:2018ime, Ivanov:2019kyh}. 
The idea is that it is not obligatory to use basis invariants in order to establish basis-independent
conditions. One can also formulate these conditions in the form of basis-independent
relations among {\em basis-covariant objects} \cite{Ivanov:2019kyh}. 
Using this approach, the basis-independent necessary and sufficient conditions 
were formulated for the usual $CP$ symmetry \cite{Nishi:2006tg} and for the $CP$ symmetry of order 4 (CP4) \cite{Ivanov:2018ime},
as well as for the simultaneous presence of the two forms of $CP$ symmetry.
With these results, the issue of explicit $CP$ conservation in 3HDMs is now settled.

\subsection{Towards basis independent recognition of other symmetries in 3HDM}

The ``success story'' above supports the idea of using basis-covariant objects of the bilinear formalism
to detect all other symmetries of 3HDMs.
This is what we accomplish in the present work for all the realizable symmetry groups, abelian and non-abelian.
The essence of our procedure is the following. We select a symmetry group, 
write the general Higgs potential invariant under it in a convenient basis, derive the bilinear-space
objects in that basis, identify their structural properties, and then establish basis-invariant criteria
which implement these features. The end result is a set of Checks which can be performed in any basis,
such that the model possesses a given symmetry group if and only if the potential passes these Checks.

The layout of the paper is as follows. In section~\ref{section-bilinear} we outline the bilinear space technique,
describe the products of the adjoint space vectors based on the $SU(3)$ invariant tensors $f_{ijk}$
and $d_{ijk}$, and then show the idea of dissecting the adjoint space with the aid of these vectors.
These tools will play the crucial role in detecting symmetries in a basis-invariant way. 
Then, in following three sections, we apply these methods to all symmetry groups
available in the 3HDM scalar sector, starting with the abelian ones, then continuing to non-abelian ones.
We then conclude with an outlook of how to use the results of this paper in phenomenological 
scans of the 3HDM parameter space.
Additional technical details and derivations are contained in Appendices.


\section{Bilinear space formalism}\label{section-bilinear}

\subsection{Orbit space}

We begin with a brief review of the bilinear formalism 
with specific application to 3HDMs \cite{Ivanov:2010ww,Maniatis:2014oza}.
We work with $N=3$ Higgs doublets $\phi_a$, $a=1,2,3$, all having the same electroweak quantum numbers.
The most general renormalizable 3HDM potential can be compactly written as
\be
V=Y_{ab} (\phi^{\dagger}_a\phi_b)+Z_{abcd} (\phi^{\dagger}_a\phi_b)(\phi^{\dagger}_c\phi_d)\,.\label{potential}
\ee
We construct the following $1+8$ gauge-invariant bilinear combinations $(r_0, r_i)$:
\be
r_0 = {1\over\sqrt{3}}\phi^{\dagger}_a\phi_a\,,\quad r_i = \phi^{\dagger}_a (t^i)_{ab}\phi_b\,,\quad
i=1,\dots,8\,.\label{bilinears}
\ee
Here, $t_i = \lambda_i/2$ are generators of the $SU(3)$ algebra satisfying
\be
[t_i,t_j] = i f_{ijk} t_k\,,\quad \text{and} \quad \{t_i,t_j\} = {1 \over 3}\delta_{ij}\id_{3} + d_{ijk} t_k\,,\label{structure-constants}
\ee
with the $SU(3)$ structure constants $f_{ijk}$ and the fully symmetric $SU(3)$ invariant tensor $d_{ijk}$.
With the usual choice of basis for the Gell-Mann matrices $\lambda_i$, these have the non-zero components
\be
f_{123} = 1\,, \quad
f_{147} = -f_{156} = f_{246} = f_{257} = f_{345} = -f_{367} = {1\over 2}\,,\quad
f_{458} = f_{678} = {\sqrt{3} \over 2}\,,
\label{tensor-fijk}
\ee
as well as
\bea
&&d_{146} = d_{157} = - d_{247} = d_{256} = {1\over 2}\,,\qquad
\phantom{-} d_{344} = d_{355} = - d_{366} = - d_{377} = {1\over 2}\,,\nonumber\\
&& d_{118} = d_{228} = d_{338} = - d_{888} = {1\over \sqrt{3}}\,,\qquad
d_{448} = d_{558} = d_{668} = d_{778} = -{1\over 2\sqrt{3}}\,.\label{tensor-dijk}
\eea
Group-theoretically, $r_0$ is an $SU(3)$ singlet and $r_i$ realizes the adjoint representation of $SU(3)$.
The coefficient in the definition of $r_0$ is not fixed by this construction. We use here the definition
borrowed from \cite{Ivanov:2010ww} but alternative normalization factors are possible \cite{Maniatis:2014oza};
the exact choice is not essential here.
In the Gell-Mann basis, the bilinears $r_i$ have the following form:
\bea
&&r_1 + i r_2 = \phi_1^\dagger \phi_2\,,\quad
r_4 + i r_5 = \phi_1^\dagger \phi_3\,,\quad
r_6 + i r_7 = \phi_2^\dagger \phi_3\,,\nonumber\\
&&
r_3 = \fr{1}{2}(\phi_1^\dagger\phi_1 - \phi_2^\dagger\phi_2)\,,\quad
r_8 = \fr{1}{2\sqrt{3}}(\phi_1^\dagger\phi_1 + \phi_2^\dagger\phi_2 - 2 \phi_3^\dagger\phi_3)\,.\label{ri-explicit}
\eea
The real vectors $r$ obtained in this way do not fill the entire real eight-dimensional space $\RR^8$ 
(the \emph{adjoint space}, whose vectors will be denoted as $x$),
but a 7D manifold in it, which is called the orbit space. The points of this space are in one-to-one correspondence
with gauge orbits within the Higgs fields space $\phi_a$.
Algebraically, the orbit space is defined by the following (in)equalities \cite{Ivanov:2010ww}:
\be
r_0 \ge 0\,, \quad r_0^2 - r_i^2 \ge 0\,,\quad
d_{ijk}r_ir_jr_k + {1\over 2\sqrt{3}}r_0(r_0^2 - 3 r_i^2) = 0\,.\label{orbit-space-r}
\ee
A basis change in the space of Higgs doublets $\phi_a \to U_{ab} \phi_b$ with $U \in SU(3)$
leaves $r_0$ unchanged and induces an $SO(8)$ rotation of the vector $r_i$.
However, not all $SO(8)$ rotations can be obtained in this way;
they must conserve, in addition, $d_{ijk}r_i r_j r_k$.

\subsection{Constructions in the adjoint space}\label{section-adjoint}

The main advantage of changing
to the bilinear space is that
the potential $V$ becomes a quadratic rather than quartic function of variables:
\be
V = M_0 r_0 + M_i r_i + \Lambda_{0}r_0^2 + L_i r_0 r_i + \Lambda_{ij}r_i r_j\,.\label{V-bilinears}
\ee
This generic expression holds for any NHDM. 
$M_0$, $\Lambda_0$, the entries of the real vectors $M$, $L$ lying in the adjoint space $\RR^{N^2-1}$,
and the $(N^2-1)\times (N^2-1)$ entries of the real symmetric matrix $\Lambda$, are all expressed in terms of the components
of the tensors $Y_{ab}$ and $Z_{abcd}$ in \eqref{potential}.

In 2HDMs, any $SO(3)$ rotation in the adjoint space can be induced by a basis change of the two Higgs doublets.
Therefore, the matrix $\Lambda$ can always be diagonalized and its eigenvectors
can always be aligned with the axes $x_1$, $x_2$, and $x_3$.
These eigenvectors as well as the vectors $M$, $L$ are covariant objects and transform
in the same way under basis changes.
Using $SO(3)$ invariant tensors $\delta_{ij}$ and $\epsilon_{ijk}$,
one can contract these vectors and obtain basis invariants.

In 3HDMs, the potential \eqref{V-bilinears} contains
two 8D vectors $M$ and $L$ and the $8 \times 8$ real symmetric matrix $\Lambda$.
The lack of the full $SO(8)$ rotational freedom 
implies that it is not guaranteed 
that $\Lambda$ can be diagonalized by a basis change.
Nevertheless, $\Lambda$ can always be expanded over its eigensystem,
and eigenvalues and eigenvectors can be found, at least numerically.

We can now formulate the main idea which was recently proposed in \cite{Ivanov:2019kyh}
and which we fully develop in the present work.
\begin{itemize}
\item[]
The basis-invariant information encoded in the eigensystem of $\Lambda$ 
and in the vectors $M$ and $L$ completely determines all physically relevant 
structural properties of the scalar sector of a 3HDM.
Although all the vectors in the adjoint space are not invariant under basis changes,
their relative orientation as well as their orientation with respect to the 
orbit space \eqref{orbit-space-r} is basis invariant.
The challenge is to extract this basis-invariant information and
to link it to the symmetry groups of 3HDMs.
\end{itemize}
The main tool which will help us overcome this challenge is to make full use
of the two additional $SU(3)$ invariant tensors $f_{ijk}$ and $d_{ijk}$ defined
in \eqref{structure-constants}. 
Given any two vectors $a$ and $b$ in the adjoint space, 
one can use these tensors to define their $f$- and $d$-products as well as 
a non-linear action on a vector:
\be
F^{(ab)}_i \equiv f_{ijk} a_j b_k\,,\quad D^{(ab)}_i \equiv \sqrt{3} d_{ijk} a_j b_k\,,
\quad D^{(aa)}_i \equiv \sqrt{3} d_{ijk} a_j a_k\,.
\label{fd-products}
\ee
These products respect group covariance: vectors $F$ and $D$ transform as adjoint $SU(3)$ representations
and, if needed, can be used in additional products.\footnote{%
Group-theoretically, vectors $F$ and $D$ represent the antisymmetric and symmetric octets
of the direct product $8 \otimes 8$ of $SU(3)$. Using appropriate projectors,
one can also extract higher-dimensional representations out of this product.
However since they reside in a different space, we do not involve them in our analysis.}

These products were first used in \cite{Nishi:2006tg} as building blocks of the basis-invariant algorithm
to detect the usual $CP$ symmetry in 3HDMs. 
For more than a decade, there were no follow-up studies. In fact, it was not broadly acknowledged
by the community that these basis-invariant conditions for explicit $CP$ conservation had been 
established in 3HDMs.
Very recently, this approach was revived and further developed in \cite{Ivanov:2018ime} where the
basis-invariant conditions for CP4 were established.
These two papers provide the complete answer to the question of 
the basis-invariant recognition of a $CP$ symmetry in 3HDMs
and the same methodology enables the detection of other symmetries possible in 3HDMs.
This is what we are going to achieve in the present paper.

\subsection{Properties of the $f$ and $d$-products}\label{fd-properties}

The vectors $F$ and $D$ defined in \eqref{fd-products} obey certain remarkable properties,
which follow from various relations among $SU(3)$-invariant tensors, see e.g. \cite{Borodulin:2017pwh}.
First, using the Jacobi identity $d_{ijk}f_{klm} + d_{jlk}f_{kim} + d_{lik}f_{kjm} = 0$, 
one observes that vectors $F^{(ab)}$ and $D^{(ab)}$ are always orthogonal: 
\bea
D^{(ab)}_k F^{(ab)}_k &=& \sqrt{3} a_{i}a_{i'}b_{j}b_{j'}d_{ijk}f_{ki'j'} =
\sqrt{3} a_{i}a_{i'}b_{j}b_{j'} (- d_{ji'k}f_{kij'} - d_{i'ik}f_{kjj'}) \nonumber\\
&=& - D^{(ab)}_k F^{(ab)}_k = 0\,.
\eea
Any of these two vectors can be zero, but not simultaneously, because their norms
satisfy
\be
|D^{(ab)}|^2 + |F^{(ab)}|^2 = \vec a^2 \vec b^2\,.\label{DF1}
\ee
For contraction of two $d$'s, one has in $SU(3)$ the following relation:
\be
d_{ijk}d_{klm} + d_{jlk}d_{kim} + d_{lik}d_{kjm} = 
{1\over 3}\left(\delta_{ij}\delta_{lm} + \delta_{il}\delta_{jm} + \delta_{im}\delta_{jl}\right)\,.
\ee
Using it, one can derive
\be
D^{(aa)} D^{(ab)} = \vec a^2\,(\vec a\vec b)\,,\quad
|D^{(aa)}|^2 = (\vec a^2)^2\,,\quad
D^{(aa)} D^{(bb)} + 2 |D^{(ab)}|^2 = \vec a^2 \vec b^2 + 2 (\vec a\vec b)^2\,.
\label{DD}
\ee
This means that the non-linear action of $d$ defined via $a \mapsto D^{(aa)}$
preserves the norm of unit vectors. 
If $a$ and $b$ are orthonormal, then $D^{(aa)}$ and $D^{(ab)}$ are orthogonal and
the absolute value of $D^{(ab)}$ can be computed from the last relation:
$|D^{(ab)}| = \sin(\varphi_{AB}/2)$, where $\varphi_{AB}$
is the angle between vectors $D^{(aa)}$ and $D^{(bb)}$.
In particular, if it happens that $D^{(aa)} = D^{(bb)}$, then $D^{(ab)} = 0$,
while if $D^{(aa)} = - D^{(bb)}$, then $|D^{(ab)}|=1$.

\subsection{Detecting subspaces}\label{detecting-subspaces}

The expressions for the tensors $f_{ijk}$ and $d_{ijk}$ make it clear that not all directions 
in the adjoint space $\RR^8$ are equivalent.
There are basis-invariant features which distinguish various subspaces of $\RR^8$ with equal dimensions.
We will see below that 3HDMs equipped with various symmetry groups
differ by the subspaces in which the vectors $M$ and $L$ and the eigenvectors of $\Lambda$ reside.
Therefore, the first key step towards our goal is to develop a set of basis-invariant checks 
which detect that (eigen)vectors belong to a subspace of $\RR^8$ with certain properties.

The checks which are described in this section and elaborated in full detail in the appendix
will be used to detect the direction $x_8$, the subspace $(x_3,x_8)$, 
various patterns of the matrix $\Lambda$ in its
orthogonal complement
\be
V_6 = (x_1,x_2,x_4,x_5,x_6,x_7)\,,\label{V6-subspace}
\ee
among others. We stress 
that these checks detect certain basis-invariant conditions. It is never
needed to actually switch to a preferred basis to perform a check.
For example, ``detecting an eigenvector in direction $x_8$''
means detecting basis-invariant conditions
which indicate that there exists a basis choice 
where that vector is aligned with $x_8$.

To illustrate the detection technique, let us consider a unit vector $a$ in the adjoint space
and compute $D^{(aa)}$. Then, one observes that 
$D^{(aa)} = - a$ if and only if there exists
a basis in which $a$ is aligned along $x_8$.

The proof follows by direct calculation.
The vectors $a$ of the adjoint space $\RR^8$ are in one-to-one correspondence
with traceless hermitian $3\times 3$ matrices $A = 2 a_i t_i$, $a_i = \Tr(At_i)$.
The hermitian matrix $A$ can always be diagonalized by a basis change.
Back in the adjoint space, this means that any vector $a$ can be brought to the $(x_3,x_8)$
subspace.
Using the explicit expressions for the components of $d_{ijk}$ given in \eqref{tensor-dijk},
which are valid in any basis, one finds that $D^{(aa)}$ also stays in the same $(x_3,x_8)$
subspace:
\be
D^{(aa)}_3 = 2 a_3 a_8\,, \quad D^{(aa)}_8 = a_3^2 - a_8^2\,, \quad |D^{(aa)}|^2 = (a_3^2 + a_8^2)^2 = 1\,.
\ee
In polar coordinates on the $(x_3,x_8)$ plane, this operation acts on the angular variable 
of $a$ as $\alpha \mapsto \pi/2 - 2\alpha$.
Hence, the three directions $\alpha = \pi/2, \pi/6$, and $5\pi/6$ are stable under this action (cf.\ \cite{Ivanov:2010ww} for more details on this construction).
The first direction corresponds to $a$ being aligned with $x_8$, 
while the other two directions can be brought to it by a basis change
(a cyclic permutation of the three doublets).
Finally, if one insists on the sign in the relation $D^{(aa)} = - a$, then the unit vector $a$
must be aligned with the positive $x_8$ direction.

The observation just made is the basis of what we call {\bf Check-(8)}: 
if there exists an eigenvector of $\Lambda$, denoted $e^{(8)}$, which satisfies  
\be
D^{(88)} = -e^{(8)}\,,\label{conditionD88}
\ee
then, in the appropriate basis, $e^{(8)}$ is along axis $x_8$, and the matrix $\Lambda$
takes the block-diagonal form with a $7\times 7$ block and a stand-alone entry $\Lambda_{88}$.
Such an eigenvector does not have to be unique. 

Next, let us find when two adjoint space vectors $a$ and $b$ can be simultaneously brought 
to the $(x_3,x_8)$ subspace. This is possible if and only if the corresponding traceless hermitian matrices $A$ and $B$ commute. Back in the adjoint space,
this is equivalent to 
\be
F^{(ab)} = 0\,.\label{f-orthogonality}
\ee
Thus, we obtain {\bf Check-(38)}: if $\Lambda$ has two orthogonal eigenvectors $a$
and $b$ which satisfy \eqref{f-orthogonality}, then there exists a basis change which brings both of them 
to the $(x_3,x_8)$ plane. The matrix $\Lambda$ takes the block-diagonal form 
with a $2\times 2$ block in this subspace and the $6\times 6$ block in its orthogonal complement 
$V_6$. Again, it is not guaranteed that such a pair of eigenvectors is unique.

One can give an alternative formulation for Check-(38) using $d$-products.
Indeed, due to Eq.~\eqref{DF1}, $f$-orthogonality implies that $|D^{(ab)}|=1$,
and then, using Eq.~\eqref{DD}, one obtains that $D^{(aa)} D^{(bb)} = -1$.
This is only possible if $D^{(aa)} = - D^{(bb)}$. One can also show the converse:
starting from $D^{(aa)} = - D^{(bb)}$ for two orthogonal eigenvectors of $\Lambda$,
one recovers Eq.~\eqref{f-orthogonality}.

Notice that passing Check-(38) does not guarantee that the two eigenvectors
are aligned with the axes $x_3$ and $x_8$. For that, one needs to require an extra condition,
and the criterion for this to happen can be summarized as
\be
D^{(aa)} = - D^{(bb)} = -a\,.\label{detecting-3and8}
\ee
We thus formulate {\bf Check-(3)(8)}:
if matrix $\Lambda$ has two eigenvectors $a$ and $b$ satisfying \eqref{detecting-3and8},
then, in the appropriate basis, $a$ is aligned with $x_8$ and $b$ is aligned with $x_3$.

The two eigenvectors emerging from Check-(3)(8) appear in it on different footing.
It becomes clear if one reformulates this Check as a two-step procedure:
first, perform Check-(8) to detect $a$, and then observe that there exists another vector $b$
such that $F^{(ab)}=0$. This second vector can only be within the subspace $(x_1,x_2,x_3)$,
and it can be aligned with $x_3$ if needed.
This procedure makes it evident that vector $b$ is not unique; there is the entire 3D subspace
which is both orthogonal and $f$-orthogonal to the vector $a$ passing Check-(8).

This observation allows us to formulate {\bf Check-(123)(8)}:
if $\Lambda$ passes Check-(8) and if, in addition, it has three other mutually orthogonal
eigenvectors $b, b', b''$ which are orthogonal and $f$-orthogonal to $e^{(8)}$
\be
F^{(b8)} = F^{(b'8)} = F^{(b''8)} = 0\,,\label{detecting-123-8}
\ee
then in an appropriate basis, $e^{8}$ is along $x_8$, while
vectors $(b,b',b'')$ span the subspace $(x_1,x_2,x_3)$.
If needed, these eigenvectors can be aligned with the axes by a basis change.
Thus, the matrix $\Lambda$ takes in this basis the block-diagonal form
with the diagonal entries $\Lambda_{11}$, $\Lambda_{22}$, $\Lambda_{33}$, $\Lambda_{88}$,
and the $4\times 4$ block in the orthogonal complement 
\be
V_4 = (x_4,x_5,x_6,x_7)\,.\label{V4}
\ee
These simple examples give an overall impression of how one can detect
subspaces in $\RR^8$ with distinct basis-invariant properties 
and ensure that $\Lambda$ has certain block-diagonal form in an appropriate basis.
In the Appendix, we further develop this technique and derive several other checks.
We also add here that, when deriving properties of certain subspaces,
one often has a choice of which vectors to use, $F$ or $D$.
Most checks below we will make use of vectors $D$, although in some cases
an equivalent formulation in terms of vectors $F$ is also possible,
in the light of the relations listed in the previous subsection.

Sparing the details presented in Appendices~\ref{appendix-subspaces} and \ref{appendix-B}, 
we give here a list of the checks for $\Lambda$, which detect various special subspaces
or patterns inside subspaces.
\begin{itemize}
\item Check-(8) detects a 1D subspace ($\Lambda$ acquires a block-diagonal form with blocks 7+1);
\item Check-(38) detects a 2D subspace (6+2);
\item Check-(3)(8) detects two 1D subspaces (6+1+1);
\item Check-(123)(8) detects four 1D subspaces (4+1+1+1+1);
\item 
Check-(123) and Check-(257) detect two inequivalent 3D subspaces (5+3), 
closely related to the $SU(2)$ and $SO(3)$ subgroups of $SU(3)$;
\item Check-(1238), or equivalently Check-(4567), detects a 4D subspace (4+4).
\end{itemize}
For $\Lambda$ matrices passing Check-(38), the 6D subspace $V_6$ 
can further split or can demonstrate special patterns.
\begin{itemize}
\item Check-(12) detects another 2D subspace (4+2+2);
\item Check-(12)(45)(67) detects all 2D subspaces (2+2+2+2);
\item Check-$\Z_3$ detects a pattern within $V_6$ characteristic for the $\Z_3$ symmetry group.
\end{itemize}
For $\Lambda$ matrices passing Check-(38) and Check-(12), 
the 4D subspace $V_4$ can still demonstrate special patterns characteristic for 
two non-equivalent implementations of $U(1)$ symmetry:
\begin{itemize}
\item Check-$U(1)_1$; 
\item Check-$U(1)_2$.
\end{itemize}

In the following sections, we will show how various symmetries groups imposed on the 3HDM scalar sector
can be detected in the basis-invariant way via these Checks.

\section{Abelian groups\label{section-abelian-groups}}

\subsection{Rephasing symmetries}

Let us first recapitulate the main features of the classification of abelian symmetry groups 
in the scalar sector of 3HDMs \cite{Ferreira:2008zy, Ivanov:2011ae}.
All abelian subgroups of $SU(3)$, in a certain basis, 
can be represented by rephasing groups.\footnote{As explained in \cite{Ivanov:2011ae},
for a proper construction, one should use abelian subgroups of $PSU(3)$ rather then $SU(3)$. In this case,
one additional abelian subgroup appears, $\Z_3 \times \Z_3$.
}
Only a few of them can be used to define models which do not possess additional accidental family symmetries. 
These groups are:
\be
\Z_2\,, \quad \Z_3\,, \quad \Z_4\,, \quad \Z_2\times \Z_2\,, \quad U(1)\,, \quad U(1)\times \Z_2\,, \quad U(1)\times U(1)\,.\label{all-abelian}
\ee
All of them are subgroups of the maximal abelian group $U(1)\times U(1)$. 
Qualitatively, the larger the symmetry group is, the fewer are the free parameters remaining in the potential,
and the tighter are the conditions one needs to impose to define the model.

The maximal abelian group $U(1)\times U(1)$ (maximal torus) is a two-parametric subgroup of $SU(3)$ 
of the following transformations:
\be
U(1)_1 = \diag(e^{i\alpha},\, e^{-i\alpha},\, 1)\,,\quad U(1)_2 = \diag(e^{-2i\beta},\, e^{i\beta},\, e^{i\beta})\,,\quad \alpha, \beta \in [0, 2\pi]\,.
\label{U1U1}
\ee
Notice that the two transformations $U(1)_1$ and $U(1)_2$ differ by their eigenvalue multiplicities. 
There is no basis change which would map any $U(1)_1$-transformation into any $U(1)_2$ transformation.
Also, notice that $Z(SU(3)) \simeq \Z_3$, the center of $SU(3)$ generated by $\diag(\omega,\omega,\omega)$ 
is located inside $U(1)_2$. 
If one wants to construct the maximal torus in $PSU(3) \simeq SU(3)/Z(SU(3))$, 
one would get the same $U(1)_1$ and $U(1)_2$, as \eqref{U1U1} but with $\beta \in [0, 2\pi/3]$.

\subsection{$U(1)\times U(1)$ 3HDM}

Let us now write the 3HDM potential symmetric under $U(1)\times U(1)$:
\be
V_0 = \sum_{a} m_{a}^2 \phi_a^\dagger \phi_a
+ \sum_{a} \lambda_{a} (\phi_a^\dagger \phi_a)^2 
+ \sum_{a < b} \left[\lambda_{ab} (\phi_a^\dagger \phi_a) (\phi_b^\dagger \phi_b) + 
\lambda'_{ab} (\phi_a^\dagger \phi_b) (\phi_b^\dagger \phi_a)\right]\,.\label{V0}
\ee
It contains 3 quadratic terms and 9 quartic terms, all with real coefficients.
The model is automatically $CP$-conserving; the $CP$ symmetry can be generated, for instance,
by the usual conjugation.

In the adjoint space, one gets scalars
\be
\Lambda_0 = {1\over 3}(\lambda_1 + \lambda_2 + \lambda_3 + \lambda_{12} + \lambda_{13} + \lambda_{23})\,,
\quad M_0 = {1\over\sqrt{3}}(m_1^2 + m_2^2 + m_3^2)\,,
\ee
the two vectors 
\bea
&&M_3 = m_1^2 - m_2^2\,, \quad M_8 = {1\over\sqrt{3}}(m_1^2 + m_2^2 - 2 m_3^2)\,,\nonumber\\
&&L_3 = {2\over\sqrt{3}}\left(\lambda_1 - \lambda_2 + {\lambda_{13} - \lambda_{23} \over 2}\right)\,,\quad
L_8 = {2\over\sqrt{3}}\left(\lambda_1 + \lambda_2 - 2 \lambda_3 + \lambda_{12} - {\lambda_{13} + \lambda_{23} \over 2}\right)\,.
\label{LMU1U1}
\eea
and
\be
\Lambda = \left(
\begin{array}{cccccccc}
 \lambda'_{12} & \cdot & \cdot & \cdot & \cdot & \cdot & \cdot & \cdot \\
 \cdot & \lambda'_{12} & \cdot & \cdot & \cdot & \cdot & \cdot & \cdot \\
 \cdot & \cdot & \Lambda_{33} & \cdot & \cdot & \cdot & \cdot & \Lambda_{38} \\
 \cdot & \cdot & \cdot & \lambda'_{13} & \cdot & \cdot & \cdot & \cdot \\
 \cdot & \cdot & \cdot & \cdot & \lambda'_{13} & \cdot & \cdot & \cdot \\
 \cdot & \cdot & \cdot & \cdot & \cdot & \lambda'_{23}& \cdot & \cdot \\
 \cdot & \cdot & \cdot & \cdot & \cdot & \cdot & \lambda'_{23} & \cdot \\
 \cdot & \cdot & \Lambda_{38} & \cdot & \cdot & \cdot & \cdot & \Lambda_{88} \\
\end{array}\label{V0-Lam}
\right)
\ee
with
\bea
&&\Lambda_{33} = \lambda_1 + \lambda_2 - \lambda_{12}\,,\quad
\Lambda_{38} = {1\over\sqrt{3}}\left(\lambda_1 - \lambda_2 - \lambda_{13} + \lambda_{23}\right)\,,\nonumber\\
&&\Lambda_{88} = {1\over 3}\left(\lambda_1 + \lambda_2 + 4 \lambda_3 + \lambda_{12} - 2\lambda_{13} - 2\lambda_{23}\right)\,.
\eea
One observes that $\Lambda$ has a generic $2\times 2$ block in the $(x_3,x_8)$ subspace,
while in the subspace $V_6$ (Eq.~ \eqref{V6-subspace}),
it has the diagonal, pairwise-degenerate structure within the subspaces $(x_1,x_2)$, $(x_4,x_5)$, and $(x_6,x_7)$.
The two vectors $M$ and $L$ have non-zero components in the $(x_3,x_8)$ subspace.

Using the results of sections~\ref{detecting-subspaces} and 
appendix \ref{splitting-V6-D}, we can easily formulate
necessary and sufficient basis-invariant conditions for the 3HDM potential to be $U(1)\times U(1)$ symmetric:
\begin{itemize}
\item
the matrix $\Lambda$ must pass Check-(38) and Check-(12)(45)(67);
\item
each pair of eigenvectors in Check-(12)(45)(67) must correspond to the same eigenvalue;
\item
the vectors $M$ and $L$ must be orthogonal to the six eigenvectors of $V_6$.
\end{itemize}

\subsection{$U(1)$ 3HDM}

Groups $U(1)_1$ and $U(1)_2$ in \eqref{U1U1} are distinct, and imposing each of them 
constrains the potential in a different way.
Imposing $U(1)_1$ leads, in addition to $V_0$ (Eq.~\eqref{V0}), to one more term:
\be
V_{U(1)_1} = \lambda_5 (\phi_1^\dagger\phi_3) (\phi_2^\dagger\phi_3) + h.c.\label{VU11}
\ee
with complex $\lambda_5$. Since $\lambda_5$ is the only complex parameter, 
one can rephase the doublets to make it real, which implies that $U(1)_1$ automatically leads 
to explicit $CP$ conservation.
In the adjoint space, the blocks of $\Lambda$ in $(x_3,x_8)$ and $(x_1,x_2)$ are unchanged, 
while within the subspace $V_4$ (Eq.~\eqref{V4}),
the block is modified by the additional term to
\be
\mtrx{
\lambda'_{13} & 0 & \Re\lambda_5 & -\Im\lambda_5 \\
0 & \lambda'_{13} & -\Im\lambda_5 & -\Re\lambda_5 \\
\Re\lambda_5 & -\Im\lambda_5 & \lambda'_{23} & 0 \\
-\Im\lambda_5 & -\Re\lambda_5 & 0 & \lambda'_{23}}\,.\label{U11-block4x4}
\ee
This pattern in $V_4$ can be detected by conditions formulated in Appendix~\ref{structures-V4}.
Thus, the necessary and sufficient basis-invariant conditions for the 3HDM potential 
to be $U(1)_1$-symmetric are:
\begin{itemize}
\item
the matrix $\Lambda$ passes Check-(38) and Check-(12);
\item
the two eigenvectors of Check-(12) correspond to the same eigenvalue;
\item
within $V_4$, $\Lambda$ passes Check-$U(1)_1$;
\item
the vectors $M$ and $L$ are orthogonal to the six eigenvectors in $V_6$.
\end{itemize}

In contrast to $U(1)_1$, $U(1)_2$ allows for several new terms in addition to $V_0$:
\be
V_{U(1)_2} = m_{23}^2 \phi_1^\dagger \phi_2 + \bar{\lambda}_5 (\phi_1^\dagger \phi_2)^2
+ (\phi_1^\dagger \phi_2) \left(\bar{\lambda}_6 \phi_1^\dagger \phi_1 + \bar{\lambda}_7 \phi_2^\dagger \phi_2 
+ \bar{\lambda}_8 \phi_3^\dagger \phi_3\right) + \bar{\lambda}'_8 (\phi_1^\dagger \phi_3)(\phi_3^\dagger \phi_2)
+ h.c.\label{VU12}
\ee
All coefficients here can be complex. Even if one sets some of them real by a basis change, several complex coefficients will remain. Thus, the $U(1)_2$-symmetric 3HDM can be explicitly $CP$ violating.

In the adjoint space, one sees that vectors $M$ and $L$ can now have unconstrained components
in the subspace $(x_1, x_2, x_3, x_8)$. 
The matrix $\Lambda$ has a block-diagonal form with two blocks $4\times 4$.
The block in the subspace $(x_1, x_2, x_3, x_8)$ is generic, and therefore its eigenvalues are unconstrained.
The block in its orthogonal complement $V_4$ shows the following pattern:
\be
\mtrx{
\lambda'_{13} & 0 & \Re\bar{\lambda}'_8 & \Im\bar{\lambda}'_8 \\
0 & \lambda'_{13} & -\Im\bar{\lambda}'_8 & \Re\bar{\lambda}'_8 \\
\Re\bar{\lambda}'_8 & -\Im\bar{\lambda}'_8 & \lambda'_{23} & 0 \\
\Im\bar{\lambda}'_8 & \Re\bar{\lambda}'_8 & 0 & \lambda'_{23}}\,,\label{U12-block4x4}
\ee
which is different from \eqref{U11-block4x4}.
Thus, the necessary and sufficient basis-invariant conditions for the 3HDM potential 
to be $U(1)_2$-symmetric are:
\begin{itemize}
\item
the matrix $\Lambda$ passes Check-(4567) described in Appendix~\ref{detecting-4D};
\item
within $V_4$, $\Lambda$ passes Check-$U(1)_2$ described in Appendix~\ref{structures-V4};
\item
the vectors $M$ and $L$ are orthogonal to the four eigenvectors in $V_4$.
\end{itemize}

\subsection{$U(1)\times \Z_2$ 3HDM}

If one keeps, out of all terms in $V_{U(1)_2}$, only $\bar{\lambda}_5 (\phi_1^\dagger \phi_2)^2 + h.c.$,
then the potential is invariant not only under $U(1)_2$ but also under the $\Z_2$ subgroup of $U(1)_1$, 
which flips the sign of $\phi_1$. 
Since we are left with only one complex coefficient, this model is explicitly $CP$ conserving.

The new term preserves the block-diagonal form of $\Lambda$ in Eq.~\eqref{V0-Lam} 
apart from the $2\times 2$ block in the $(x_1, x_2)$ subspace. 
This block becomes generic, so that the eigenvalue degeneracy is lifted.
Thus, the basis-invariant conditions for the $U(1)\times \Z_2$ 3HDM are the same as for 
$U(1)\times U(1)$ 3HDM with only this condition relaxed.

\subsection{$\Z_2 \times \Z_2$ 3HDM}\label{section-Z2Z2}

Restricting the previous case to the discrete subgroup of arbitrary sign flips, one obtains the famous Weinberg model
with the symmetry group $\Z_2 \times \Z_2$  \cite{Weinberg:1976hu}.
The Higgs potential contains, in addition to $V_0$, the following three terms:
\be
V_{\Z_2 \times \Z_2} = \bar{\lambda}_{12} (\phi_1^\dagger \phi_2)^2 + \bar{\lambda}_{23} (\phi_2^\dagger \phi_3)^2 
+ \bar{\lambda}_{31} (\phi_3^\dagger \phi_1)^2 + h.c.,\label{Z2Z2}
\ee
where all coefficients can be complex. If $\Im(\bar{\lambda}_{12}\bar{\lambda}_{23}\bar{\lambda}_{31}) \not = 0$,
then it is impossible to make all coefficients real by any basis change, and the model is explicitly $CP$ violating.\footnote{%
Even with complex coefficients, there remains the possibility of a generalized $CP$ symmetry 
which does not commute with $\Z_2 \times \Z_2$. This case is treated in section~\ref{section-exoticCP}.}
If it is real, then the model is explicitly $CP$ conserving and is known as Branco's model \cite{Branco:1979pv,Branco:1980sz}.

In the adjoint space, the generic form within the subspace $(x_3,x_8)$ is unchanged, while in $V_6$
the new terms \eqref{Z2Z2} with generic complex 
$\bar{\lambda}_{12}$, $\bar{\lambda}_{23}$, and $\bar{\lambda}_{31}$,
drive the completely diagonal $\Lambda$ of Eq.~\eqref{V0-Lam} 
into a block-diagonal form with three $2\times 2$ blocks within subspaces $(x_1,x_2)$, $(x_4,x_5)$, and $(x_6,x_7)$. 
All these blocks are generic, so that the eigenvalues are not constrained.
Using the rephasing freedom, one can diagonalize at least two of the three blocks.
If the third one also gets diagonalized, we have the explicitly $CP$-conserving case (Branco's model),
if not, we have the explicitly $CP$-violating case (Weinberg's model).

The necessary and sufficient basis-invariant conditions for the $\Z_2\times \Z_2$-symmetric 3HDM 
are given by the simplified version of the $U(1)\times U(1)$ case: 
\begin{itemize}
\item
the matrix $\Lambda$ passes Check-(38) and Check-(12)(45)(67);
\item
the vectors $M$ and $L$ are orthogonal to the six eigenvectors in $V_6$.
\end{itemize}
Explicit $CP$ conservation, that is, whether this is Weinberg's or Branco's model,
can be detected by Check-(257) described in Appendix~\ref{detecting-3D} 
and first derived in \cite{Nishi:2006tg}.

\subsection{$\Z_4$ 3HDM \label{sec:Z4}}

The $\Z_4$-symmetric 3HDM can only arise as a particular case of the $U(1)_1$ 3HDM.
The $\Z_4$-symmetric potential contains, in addition to $V_0$, two extra terms:
\be
V_{\Z_4} = \lambda_5 (\phi_1^\dagger\phi_3) (\phi_2^\dagger\phi_3) + \bar{\lambda}_{12} (\phi_1^\dagger \phi_2)^2 + h.c.\label{VZ4}
\ee
Since there are only two complex coefficients, they can be made real via rephasing, and the model is explicitly $CP$ conserving.
The matrix $\Lambda$ has the familiar features: a generic block in the subspace $(x_3,x_8)$, 
a generic block in the subspace $(x_1,x_2)$,
and the block-diagonal structure \eqref{U11-block4x4} in $V_4$.
The basis-invariant conditions are the same as for $U(1)_1$, 
with the removal of the condition of the eigenvalue degeneracy 
within the subspace $(x_1,x_2)$, i.e.
\begin{itemize}
\item
the matrix $\Lambda$ passes Check-(38) and Check-(12);
\item
within $V_4$, $\Lambda$ passes Check-$U(1)_1$;
\item
the vectors $M$ and $L$ are orthogonal to the six eigenvectors in $V_6$.
\end{itemize}

\subsection{$\Z_3$ 3HDM}

The $\Z_3$-symmetric 3HDM can also only arise as a particular case of the $U(1)_1$ 3HDM.
Its potential contains, in addition to $V_0$, three extra terms:
\be
V_{\Z_3} = \lambda_5 (\phi_1^\dagger\phi_3) (\phi_2^\dagger\phi_3) + \lambda_6 (\phi_2^\dagger\phi_1) (\phi_3^\dagger\phi_1)
+ \lambda_7 (\phi_3^\dagger\phi_2) (\phi_1^\dagger\phi_2) + h.c., \label{VZ3}  
\ee
where all coefficients can be complex. 
Even if one makes two of them real (for example $\lambda_6$ and $\lambda_7$), the other (e.g. $\lambda_5$) can still be complex, 
thus the possibility of explicit $CP$ violation remains.

The matrix $\Lambda$ still has a generic block in $(x_3,x_8)$, while within $V_6$ it takes the following form:
\be
\left(
\begin{array}{cccccc}
 \lambda'_{12} & 0 & \Re\lambda_6 & \Im\lambda_6 & \Re\lambda_7 & \Im\lambda_7  \\
 0 & \lambda'_{12} & \Im\lambda_6  & -\Re\lambda_6 & -\Im\lambda_7 & \Re\lambda_7 \\
 \Re\lambda_6 & \Im\lambda_6  & \lambda'_{13} & 0 & \Re\lambda_5 & -\Im\lambda_5 \\
 \Im\lambda_6  & -\Re\lambda_6 & 0 & \lambda'_{13} & -\Im\lambda_5 & -\Re\lambda_5 \\
 \Re\lambda_7 & -\Im\lambda_7 & \Re\lambda_5 & -\Im\lambda_5 & \lambda'_{23} & 0 \\
 \Im\lambda_7 & \Re\lambda_7 & -\Im\lambda_5 & -\Re\lambda_5 & 0 & \lambda'_{23} \\
\end{array}\label{Z3-block6x6}
\right)
\ee
This matrix has three twice-degenerate eigenvalues.
In appendix~\ref{Z3-pattern}, we prove that this pattern emerges if and only if 
all three pairs of eigenvectors corresponding to the same eigenvalue pass Check-$\Z_3$.
Therefore, the basis-invariant necessary and sufficient conditions for $\Z_3$-symmetric 3HDM
are:
\begin{itemize}
\item
the matrix $\Lambda$ passes Check-(38);
\item
the six eigenvalues of $\Lambda$ within $V_6$ display $2+2+2$ degeneracy,
and each pair of the eigenvectors passes Check-$\Z_3$;
\item
the vectors $M$ and $L$ are orthogonal to the six eigenvectors in $V_6$.
\end{itemize}
Explicit $CP$ conservation within $Z_3$ 3HDM implies that, in a certain basis, all coefficients
are real. The $6\times 6$ block then splits into two $3\times 3$ blocks, which are closed under 
the $f$-product, so that this feature can be detected by Check-(257). 

\subsection{$\Z_2$ 3HDM}

Finally, the smallest symmetry group one can impose is $\Z_2$ generated, for example, 
by the sign flip of doublet $\phi_3$.
In the adjoint space, the only feature one observes is that $\Lambda$ splits into
two $4\times 4$ blocks: one in the $(x_1, x_2, x_3, x_8)$ subspace and the other in $V_4$.
The structure of each block is unconstrained. 
In Section~\ref{detecting-4D} we formulated Check-(1238)
which detects exactly this splitting of $\Lambda$. It must be accompanied with the requirement
that vectors $M$ and $L$ are orthogonal to the eigenvectors from $V_4$.

In summary, in this section we gave basis-invariant conditions for each rephasing symmetry group
in 3HDM, starting from the largest one $U(1) \times U(1)$ and then descending to its subgroups.
As the symmetry is reduced, we see that qualitatively the conditions are gradually relaxed.


\section{Groups with 2D representations}

\subsection{$U(2)$-symmetric 3HDM}

We now move to the symmetry groups with two-dimensional irreducible representations.
As before, we begin with the largest subgroup of $SU(3)$ with 2D irreducible representation,
$U(2) \simeq SU(2)\times U(1)$.
In the basis where $SU(2)$ transformations act non-trivially on $\phi_1$, $\phi_2$
and $U(1)$ transformations are of the type $U(1)_2$,
the potential takes the form
\bea
V_{U(2)} &=& m_1^2 (\phi_1^\dagger \phi_1 + \phi_2^\dagger \phi_2) + m_3^2 \phi_3^\dagger \phi_3
+ \lambda_1 (\phi_1^\dagger \phi_1 + \phi_2^\dagger \phi_2)^2 + \lambda_3 (\phi_3^\dagger \phi_3)^2 \nonumber\\
&&
+ \lambda_{12}' [|\phi_1^\dagger \phi_2|^2-(\phi_1^\dagger \phi_1)(\phi_2^\dagger \phi_2)]
+ \lambda_{13} (\phi_1^\dagger \phi_1 + \phi_2^\dagger \phi_2) (\phi_3^\dagger \phi_3) + 
\lambda'_{13} (|\phi_1^\dagger \phi_3|^2 + |\phi_2^\dagger \phi_3|^2)\,,\label{VU2}
\eea
which is the $U(1)\times U(1)$ potential \eqref{V0} with the additional constraints
\be
m_1^2 = m_2^2\,,\quad \lambda_1 = \lambda_2\,, 
\quad\lambda_{13} = \lambda_{23}\,,\quad \lambda_{13}' = \lambda_{23}'\,,
\label{SO2-constraints}
\ee
and
\be
\lambda_{12} = 2\lambda_1 - \lambda_{12}'\,.
\label{extra-SU2-coinstraint}
\ee
In the adjoint space, one sees that the vectors $M$ and $L$, in this basis, are along axis $x_8$.
The only off-diagonal element of $\Lambda$ in \eqref{V0-Lam} is now zero, $\Lambda_{38}=0$,
so that $\Lambda$ becomes diagonal with the following unit blocks:
\be
\Lambda=
\left(
\begin{array}{ccc}
\lambda_{12}' \id_3 & \cdot &  \cdot \\
\cdot  & \lambda_{13}' \id_4 &  \cdot \\
\cdot  &  \cdot & \Lambda_{8} \\
\end{array}
\right)
\label{su(2)-Lam}
\ee
and $\Lambda_8 = 4(\lambda_1 + \lambda_3 - \lambda_{13})/3 - \lambda_{12}'/3$.
The converse is also true: if $L, M$ are parallel to $x_8$ and $\Lambda$ exhibits this pattern, 
then the potential is invariant under $U(2)$ symmetry.

To determine the basis-invariant conditions for the $U(2)$ symmetry to be present,
we first need to detect the special direction $x_8$.
This is done by Check-(8) described in section~\ref{detecting-subspaces}:
if there exists an eigenvector $e^{(8)}$ satisfying \eqref{conditionD88},
then in the appropriate basis it can be aligned with the positive direction of axis $x_8$.
We also require that $L$ and $M$ are aligned in the same direction.
Next, one must observe that the eigenvalues of $\Lambda$ display the degeneracy pattern $3+4+1$,
with the non-degenerate eigenvalue corresponding to $e^{(8)}$.
Moreover, the eigenvectors corresponding to the triple-degenerate eigenvalue must 
pass Check-(123) described in Appendix~\ref{detecting-3D}. If all these conditions are satisfied, 
the model has the $U(2)$ symmetry.

\subsection{$O(2)$-symmetric 3HDM}

When going from $SU(2)\times U(1)$ to smaller groups with 2D irreducible representations,
one first notices that imposing $SU(2)$ alone automatically leads to an accidental $U(1)$, bringing one back to the previous case.
Thus, we consider next the symmetry group $O(2) \simeq SO(2)\rtimes \Z_2$.
When describing $SO(2)$ transformations, it is convenient to work in the basis where they are given either
by orthogonal rotations in the $(\phi_1,\phi_2)$ subspace or by rephasing transformations from $U(1)_1$.
In the former case, the extra $\Z_2$ can be generated by a reflection with respect to 
any direction in this subspace:
\be
\hat{b}_2 = \mmmatrix{c_\delta}{s_\delta}{0}{s_\delta}{-c_\delta}{0}{0}{0}{1}\,,\label{b2-real}
\ee
with $c_\delta = \cos\delta$ and $s_\delta = \sin\delta$, angle $\delta$ being a free parameter,
while in the latter case the generator can be the transformation $b_2$
\be
b_2 = \mmmatrix{0}{e^{i\delta}}{0}{e^{-i\delta}}{0}{0}{0}{0}{1}\,.\label{b2}
\ee
In the real $O(2)$ basis, the most general potential compatible with this symmetry contains,
in addition to Eq.~\eqref{VU2}, the following terms:
\be
V_{O(2)} = \bar\lambda_{12} (\Im \phi_1^\dagger\phi_2)^2 + 
\left\{
{\bar\lambda_{13} \over 2}\left[(\phi_1^\dagger \phi_3)^2 + (\phi_2^\dagger \phi_3)^2\right] 
+ h.c. \right\}\,.\label{VO2}
\ee
In the adjoint space, the matrix $\Lambda$ takes the following form:
\be
\Lambda_{SO(2)\rtimes\Z_2}=
\left(
\begin{array}{cccccccc}
\lambda_{12}' & \cdot & \cdot & \cdot & \cdot & \cdot & \cdot & \cdot \\
 \cdot & \lambda_{12}'+\bar\lambda_{12} & \cdot & \cdot & \cdot & \cdot & \cdot & \cdot \\
 \cdot & \cdot & \lambda_{12}' & \cdot & \cdot & \cdot & \cdot & \cdot \\
 \cdot & \cdot & \cdot & \lambda_{13}'+\Re\bar\lambda_{13} & -\Im\bar\lambda_{13} & \cdot &  & \cdot \\
 \cdot & \cdot & \cdot & -\Im\bar\lambda_{13} & \lambda_{13}'-\Re\bar\lambda_{13} & \cdot & \cdot & \cdot \\
 \cdot & \cdot & \cdot & \cdot & \cdot & \lambda_{13}'+\Re\bar\lambda_{13} & -\Im\bar\lambda_{13} & \cdot \\
 \cdot & \cdot & \cdot &  & \cdot & -\Im\bar\lambda_{13} & \lambda_{13}'-\Re\bar\lambda_{13} & \cdot \\
 \cdot & \cdot & \cdot & \cdot & \cdot & \cdot & \cdot & \Lambda_{8} \\
\end{array}
\right)\,.\label{SO2Z2-Lam}
\ee
In the rephasing basis, one takes $V_0$ as in \eqref{V0}, applies the conditions \eqref{SO2-constraints},
and adds the $U(1)_1$-symmetric terms \eqref{VU11} without any constraint on $\lambda_5$.
The resulting matrix $\Lambda$ acquires a slightly different form:
\be
\Lambda_{U(1)\rtimes\Z_2}=
\left(
\begin{array}{cccccccc}
\lambda_{12}' & \cdot & \cdot & \cdot & \cdot & \cdot & \cdot & \cdot \\
 \cdot & \lambda_{12}' & \cdot & \cdot & \cdot & \cdot & \cdot & \cdot \\
 \cdot & \cdot & 2\lambda_1 - \lambda_{12} & \cdot & \cdot & \cdot & \cdot & \cdot \\
 \cdot & \cdot & \cdot & \lambda_{13}' & \cdot & \Re\lambda_5 & -\Im\lambda_5 & \cdot \\
 \cdot & \cdot & \cdot & \cdot & \lambda_{13}' & -\Im\lambda_5 & -\Re\lambda_5 & \cdot \\
 \cdot & \cdot & \cdot & \Re\lambda_5 & -\Im\lambda_5 & \lambda_{13}' & \cdot & \cdot \\
 \cdot & \cdot & \cdot & -\Im\lambda_5 & -\Re\lambda_5 & \cdot & \lambda_{13}' & \cdot \\
 \cdot & \cdot & \cdot & \cdot & \cdot & \cdot & \cdot & \Lambda_{8} \\
\end{array}\label{U1Z2-Lam}
\right)\,.
\ee
In both cases one observes that the eigenvalue degeneracy pattern becomes $1+2+2+2+1$,
where the non-degenerate eigenvalues can only correspond to $x_8$
and an eigenvector in the subspace $(x_1,x_2,x_3)$.

To detect the presence of this symmetry group in a basis invariant way, 
we first detect the eigenvector $e^{(8)}$ via Check-(8)
and then the three eigenvectors in the subspace $(x_1,x_2,x_3)$ via Check-(123)(8),
described in section~\ref{detecting-subspaces}.
Next, one checks that two among the three eigenvalues within $(x_1,x_2,x_3)$ are degenerate,
which singles out the corresponding subspace $V_2$.
The exact choice depends on the basis choice; 
the two forms of $\Lambda$ in \eqref{SO2Z2-Lam} and \eqref{U1Z2-Lam} correspond to two such choices.

With these conditions, one knows that $\Lambda$ has a separate $4\times 4$ block in $V_4$
with two twice degenerate eigenvalues and one needs to establish its structure. 
Applying the methods described in Appendix~\ref{structures-V4} to any of the above two forms of $\Lambda$,
one can establish the following basis-invariant conditions.
Take a pair of eigenvectors $a$ and $b$ corresponding to the same eigenvalue.
Then they satisfy
\be
D^{(aa)} + D^{(bb)}= -e^{(8)}\,,\quad
D^{(aa)} - D^{(bb)} \in V_2\,,\quad D^{(ab)} \in V_2\,.
\label{condition-O2}
\ee
Thus, the basis-invariant algorithm for detecting an $O(2)$ symmetry in 3HDM is:
\begin{itemize}
\item
verify that $\Lambda$ passes Check-(8) and Check-(123)(8);
\item
check that at least two of the eigenvectors from the subspace $(x_1,x_2,x_3)$  correspond to the same eigenvalue;
\item
check that the remaining four eigenvectors from $V_4$ also correspond to two twice degenerate eigenvalues,
and the eigenvectors in each pair satisfy \eqref{condition-O2}.
\item
check that $L$ and $M$ are aligned with $e^{(8)}$.
\end{itemize}

\subsection{$D_4$-symmetric 3HDM}

If one starts with the $\Z_4$ symmetric model given by $V_0$ in \eqref{V0} and $V_{\Z_4}$ in \eqref{VZ4}
and imposes the conditions \eqref{SO2-constraints},
then the potential acquires yet another symmetry of order 2 given by \eqref{b2}.
No other conditions on parameters $\lambda_5$ and $\bar{\lambda}_{12}$ are needed.
The total family symmetry group is then $D_4 \simeq \Z_4 \rtimes \Z_2$, on top of which
one also has a $CP$ symmetry.
The basis-invariant algorithm for detecting this symmetry can be formulated as:
\begin{itemize}
\item
the matrix $\Lambda$ passes Check-(3)(8) and Check-(12);
\item
within $V_4$, $\Lambda$ passes Check-$U(1)_1$;
\item
the vectors $M$ and $L$ are aligned with $e^{(8)}$.
\end{itemize}

\subsection{$S_3$-symmetric 3HDM}

To construct an $S_3$-invariant 3HDM, 
one starts with the $\Z_3$-symmetric case with the potential $V_0$ in \eqref{V0}
and $V_{\Z_3}$ in \eqref{VZ3}, and imposes an additional symmetry $b_2$ (Eq.~\eqref{b2}).
As before, one obtains the same constraints \eqref{SO2-constraints} as well as the new constraint 
on the $\Z_3$-symmetric parameters:
\be
|\lambda_6| = |\lambda_7|\,.\label{S3extra}
\ee
Coefficients $\lambda_5$, $\lambda_6$, and $\lambda_7$ can still be complex with arbitrary phases,
as for any phase choice for $\lambda_6$ and $\lambda_7$, there exists a parameter $\delta$ in \eqref{b2} 
such that $b_2$ is indeed a symmetry of the potential.

In the adjoint space, we see a picture similar to the previous case.
The subspace $(x_3,x_8)$ splits into separate $x_3$ and $x_8$
subspaces, and the matrix $\Lambda$ acquires two eigenvectors along these directions, $e^{(3)}$ and $e^{(8)}$. 
The vectors $L$ and $M$ must be aligned with $e^{(8)}$.
The $6\times 6$ block of $\Lambda$ within the subspace $V_6$ keeps its form \eqref{Z3-block6x6}
but it is now constrained by the relation~\eqref{S3extra}.

We find that the shortest way to implement it in the basis-invariant way is 
to calculate vectors 
\be
K_i = d_{ijk}\Lambda_{jk}\,, \quad K^{(2)}_i = d_{ijk}(\Lambda^2)_{jk}\,,\label{KK2}
\ee
and require them to be aligned with $x_8$.
Starting from \eqref{Z3-block6x6}, one finds that the only new conditions arise from their $x_3$ components:
\bea
K_3 &=& \lambda_{13}' - \lambda_{23}' = 0\,,\nonumber\\
K^{(2)}_3 &=& \lambda_{13}^{\prime 2} + |\lambda_5|^2 + |\lambda_6|^2 - 
\left(\lambda_{23}^{\prime 2} + |\lambda_5|^2 + |\lambda_7|^2 \right)=0\,,
\eea
from which one immediately recovers \eqref{S3extra}.

In summary, the basis-invariant algorithm for $S_3$-symmetric 3HDM is:
\begin{itemize}
\item
the matrix $\Lambda$ passes Check-(3)(8);
\item
the six eigenvalues of $\Lambda$ within $V_6$ display the $2+2+2$ degeneracy,
and each pair of the eigenvectors passes Check-$\Z_3$;
\item
check that the four vectors $L$, $M$, $K$ and $K^{(2)}$ are aligned with $e^{(8)}$.
\end{itemize}
In general, the $S_3$ 3HDM can be explicitly $CP$-violating.
If one wishes to check if $CP$ is explicitly conserved, one needs to perform the same Check-(257)
which was discussed before. 

\subsection{Exotic $CP$ situations}\label{section-exoticCP}

Finally, there are two situations in which one starts with abelian Higgs family symmetry groups
but implements in addition a $CP$ symmetry in such a way that the resulting symmetry group
has 2D irreducible representation.

The first case is the 3HDM invariant under CP4.
This model was proposed in \cite{Ivanov:2015mwl} and the basis-invariant algorithm
for detecting CP4 was presented in \cite{Ivanov:2018ime}.
Formulated in the language of the present paper, this algorithm proceeds as follows, using the vectors in \eqref{KK2}:
\begin{itemize}
\item
the matrix $\Lambda$ passes Check-(8) and Check-(123)(8);
\item
the four vectors $L$, $M$, $K$ and $K^{(2)}$ are aligned with $e^{(8)}$.
\end{itemize}

The second case is the unusual realization of $CP$ symmetric $\Z_2\times \Z_2$ model,
when the $CP$ symmetry is of order 2 but it does not commute with the $\Z_2\times \Z_2$ family symmetry group.
Group-theoretically, the symmetry content is described by $(\Z_2\times \Z_2)\rtimes \Z_2^{(CP)}$ 
where the extra $\Z_2^{(CP)}$ 
is a generalized $CP$ symmetry which acts on $\Z_2\times \Z_2$ by transposing its generators $a_1$ and $a_2$:
$(CP)^{-1} a_1 CP = a_2$. This group can also be presented as generated by an order-4 $CP$ transformation
$a_1 CP$ and the usual $CP$ transformation, which do not commute. 
This model represents, therefore, a more constrained version of CP4 3HDM;
we refer to \cite{Ivanov:2018ime} for a basis-invariant strategy of detecting it.


\section{Groups with 3D representations}

\subsection{$SU(3)$-symmetric 3HDM}

Moving to symmetry groups with irreducible triplet representations, 
we begin with the largest group available, $SU(3)$.
The $SU(3)$-symmetric 3HDM has only three terms in the scalar potential:
\bea
V_{SU(3)} &=& m^2 (\phi_1^\dagger\phi_1 + \phi_2^\dagger\phi_2 + \phi_3^\dagger\phi_3)
+ \lambda (\phi_1^\dagger\phi_1 + \phi_2^\dagger\phi_2 + \phi_3^\dagger\phi_3)^2 \nonumber\\
&& \hspace{-5mm}+ \lambda' \left[ |\phi_1^\dagger \phi_2|^2 + |\phi_2^\dagger \phi_3|^2 + |\phi_3^\dagger \phi_1|^2
- (\phi_1^\dagger\phi_1)(\phi_2^\dagger\phi_2) - (\phi_2^\dagger\phi_2)(\phi_3^\dagger\phi_3) 
- (\phi_3^\dagger\phi_3)(\phi_1^\dagger\phi_1)\right].
\label{VSU3}
\eea
The second line of Eq.~\eqref{VSU3} 
represents $(\sum_{i=1}^8 r_i^2) -r_0^2$, which is a non-positive quantity. 
Thus, in the adjoint space, this potential
is characterized by vectors $L = 0$ and $M=0$ and $\Lambda = \lambda' \id_8$,
which is invariant under all $SO(8)$ rotations. Clearly, the potential will have this form in
any basis, which will be immediately recognized. Still, we can formulate
the basis-invariant condition for the $SU(3)$ symmetry as absence of any vector and the full degeneracy among
the eigenvalues of $\Lambda$.

\subsection{$SO(3)$-symmetric 3HDM}

The next possibility is to impose the $SO(3)$ subgroup of $SU(3)$.
In the basis where the $SO(3)$ generators are $t_2$, $t_5$, $t_7$, the rotations 
in the space of doublets $\phi_a$ are purely real.
Looking into how the bilinear combinations $\phi_a^\dagger \phi_b$
transform under $SO(3)$, one sees that the real symmetric combinations form the 5-plet 
of $SO(3)$ and the imaginary antisymmetric combinations form a triplet.
Therefore, back in the adjoint space, $SO(3)$ transformations do not mix 
the subspaces $V_+ = (x_1,x_3,x_4,x_6,x_8)$ and $V_- = (x_2,x_5,x_7)$.
Thus, the matrix $\Lambda$ can now be written as 
$\Lambda_1 \id_5 + \Lambda_2 \id_3$, with the two distinct eigenvalues
$\Lambda_1$ and $\Lambda_2$ corresponding to $V_+$ and $V_-$ respectively.

The basis-invariant detection of the $SO(3)$ symmetry consists in checking that
vectors $L$ and $M$ are absent, detecting the $5+3$ degeneracy pattern of the eigenvalues,
and finally verifying that the eigenvectors corresponding to the triple degenerate eigenvalue
satisfy Check-(257) described in Appendix~\ref{detecting-3D}.

\subsection{$A_4$ and $S_4$-symmetric 3HDMs}

Next, we pass to the discrete groups with irreducible triplet representation
which can arise in the scalar sector of 3HDM.
Two of them can be obtained as extensions of the $\Z_2 \times \Z_2$ group
by the permutation symmetries of three of its generators:
$A_4 \simeq (\Z_2 \times \Z_2)\rtimes \Z_3$ and $S_4 \simeq (\Z_2 \times \Z_2)\rtimes S_3$.
In the basis where $\Z_2 \times \Z_2$ is given by the sign flips of individual doublets,
the $A_4$-symmetric potential is written as a constrained version of \eqref{V0} and \eqref{Z2Z2}:
\bea
V_{A_4}&=&m^2\left(\phi_1^{\dagger}\phi_1+\phi_2^{\dagger}\phi_2+\phi_3^{\dagger}\phi_3\right)
+\lambda\left(\phi_1^{\dagger}\phi_1+\phi_2^{\dagger}\phi_2+\phi_3^{\dagger}\phi_3\right)^2\label{A4}\\
&&+\lambda_{38}\left[(\phi_1^{\dagger}\phi_1)^2+(\phi_2^{\dagger}\phi_2)^2+(\phi_3^{\dagger}\phi_3)^2-(\phi_1^{\dagger}\phi_1)(\phi_2^{\dagger}\phi_2)-(\phi_2^{\dagger}\phi_2)(\phi_3^{\dagger}\phi_3)-(\phi_3^{\dagger}\phi_3)(\phi_1^{\dagger}\phi_1)\right]\nonumber\\
&& + \lambda' \left( |\phi_1^\dagger \phi_2|^2 + |\phi_2^\dagger \phi_3|^2 + |\phi_3^\dagger \phi_1|^2\right)
+\left\{\bar{\lambda}_{12} (\phi_1^\dagger \phi_2)^2 + \bar{\lambda}_{23} (\phi_2^\dagger \phi_3)^2 
+ \bar{\lambda}_{31} (\phi_3^\dagger \phi_1)^2 + h.c.\right\}.\nonumber
\eea
Here, the parameters $\bar{\lambda}_{12}$, $\bar{\lambda}_{23}$, and $\bar{\lambda}_{31}$
can be complex with arbitrary phases but equal absolute values:
\be
|\bar{\lambda}_{12}|  = |\bar{\lambda}_{23}| = |\bar{\lambda}_{31}| \equiv \bar\lambda\,.
\ee
If these conditions are satisfied, then the potential \eqref{A4} possesses the $A_4$-symmetry,
in which the $\Z_3$ generator is given by cyclic permutations of the doublets accompanied with
suitable phase factors.
If, in addition, $\Im(\bar{\lambda}_{12}\bar{\lambda}_{23}\bar{\lambda}_{31}) =0$, the symmetry group
enlarges to $S_4$. Indeed, one can switch to the basis
where these three coefficients are real, and the potential becomes symmetric
under any (not just cyclic) permutations of the three doublets.
Notice that in either case, the model is explicitly $CP$-conserving.

In the adjoint space, one notices that $L = 0$ and $M=0$,
while the matrix $\Lambda$ takes, just as in the $\Z_2\times \Z_2$ case,
the block-diagonal form with blocks in the subspaces $(x_3,x_8)$,
$(x_1,x_2)$, $(x_4,x_5)$, and $(x_6,x_7)$.
However, the $(x_3,x_8)$ block is now simply $3\lambda_{38} \id_2$,
while the other three $2\times 2$ blocks within $V_6$ have identical pairs of eigenvalues
$\lambda' \pm 2\bar\lambda$ but arbitrarily oriented eigenvectors.
For the $S_4$-symmetric case, their orientation is correlated, though, 
and in a certain basis all eigenvectors in $V_6$ can be aligned with the axes, which renders
the matrix $\Lambda$ diagonal.
In either case, one observes the eigenvalue degeneracy pattern $2+3+3$.
Notice also that by setting, in addition, $3\lambda_{38} = \lambda' + 2\bar\lambda$, one would recover 
the $SO(3)$-symmetric case. 

The basis-invariant algorithm for detection of the $A_4$ symmetry is:
\begin{itemize}
\item
verify that the matrix $\Lambda$ passes Check-(38) with degenerate eigenvalues; 
\item
verify that $\Lambda$ passes Check-(12)(45)(67) and displays three identical pairs of eigenvalues;
\item
the vectors $L$ and $M$ are absent.
\end{itemize}
In order to detect the $S_4$ symmetry, one additionally requires that one of the triplets of $V_6$ eigenvectors sharing the same eigenvalue is closed under the action of $d$-product.

\subsection{$\Delta(54)$ and $\Sigma(36)$-symmetric 3HDM}

The symmetry group $\Delta(27) \subset SU(3)$ is generated by two order-3 transformations,
which are traditionally chosen to be rephasing transformations $\mathrm{diag}(\omega, \omega^2, 1)$
and cyclic permutations which can be accompanied by rephasings.\footnote{The commutator 
of the two generators of $\Delta(27)$ lies in the center of $SU(3)$. 
Therefore, if viewed as a subgroup of $PSU(3) \simeq SU(3)/Z(SU(3))$, 
it corresponds to the abelian group $\Z_3\times \Z_3$.
}
It turns out that $\Delta(27)$-symmetric 3HDM automatically acquires an accidental $\Z_2$ symmetry
which makes the total symmetry group of the model $\Delta(54)$.

The general $\Delta(54)$-symmetric 3HDM potential has the form similar to \eqref{A4}
but with the different last bracket:
\bea
V_{\Delta(54)}&=&m^2\left(\phi_1^{\dagger}\phi_1+\phi_2^{\dagger}\phi_2+\phi_3^{\dagger}\phi_3\right)
+\lambda\left(\phi_1^{\dagger}\phi_1+\phi_2^{\dagger}\phi_2+\phi_3^{\dagger}\phi_3\right)^2\label{Delta54}\\
&&+\lambda_{38}\left[(\phi_1^{\dagger}\phi_1)^2+(\phi_2^{\dagger}\phi_2)^2+(\phi_3^{\dagger}\phi_3)^2-(\phi_1^{\dagger}\phi_1)(\phi_2^{\dagger}\phi_2)-(\phi_2^{\dagger}\phi_2)(\phi_3^{\dagger}\phi_3)-(\phi_3^{\dagger}\phi_3)(\phi_1^{\dagger}\phi_1)\right]\nonumber\\
&& + \lambda' \left( |\phi_1^\dagger \phi_2|^2 + |\phi_2^\dagger \phi_3|^2 + |\phi_3^\dagger \phi_1|^2\right)\nonumber\\
&&+ \left\{\lambda_5 (\phi_1^\dagger\phi_3) (\phi_2^\dagger\phi_3) + \lambda_6 (\phi_2^\dagger\phi_1) (\phi_3^\dagger\phi_1)
+ \lambda_7 (\phi_3^\dagger\phi_2) (\phi_1^\dagger\phi_2) + h.c.\right\}.\nonumber  
\eea
Just like in the $\Z_3$-symmetric case, the coefficients $\lambda_5$, $\lambda_6$, and $\lambda_7$
can be complex, but, in order for the potential to be invariant under cyclic permutations,
they must have the same absolute values:
\be
|\lambda_5| = |\lambda_6| = |\lambda_7|\,.\label{conditions-delta54}
\ee
One can perform rephasing transformations to set these three parameters equal to $\bar\lambda$,
where $\bar\lambda^3 = \lambda_5\lambda_6\lambda_7$.
Additionally, if these parameters satisfy $\Im(\lambda_5\lambda_6\lambda_7) = 0$, 
then there exists a basis in which they all are real, up to powers of $\omega$, 
and the model is explicitly $CP$-conserving. 

In the adjoint space, one observes the absence of vectors $M$ and $L$, and for $\Lambda$,
the simple structure $3\lambda_{38}\id_2$ in the $(x_3,x_8)$ block, 
and the residual $6 \times 6$ block in $V_6$ which has the same structure
as \eqref{Z3-block6x6} but with equal diagonal elements and with the off-diagonal
elements satisfying the conditions \eqref{conditions-delta54}.
The eigenvalues of this $6\times 6$ block exhibit the $2+2+2$ degeneracy pattern and are equal to
\be
\lambda' + 2 \Re\bar\lambda\,, \quad \lambda' + 2 \Re(\omega \bar\lambda)\,, \quad 
\lambda' + 2 \Re(\omega^2 \bar\lambda)\,.
\ee
In the $CP$-conserving case, two of the three real parts coincide,
and the degeneracy pattern is promoted to $2+4$.

The eigenvectors of $\Lambda$ within $V_6$ can also be found explicitly:
\bea
a = {1\over\sqrt{3}}(1, 0, 1, 0, 1, 0)\,, && 
b = {1\over\sqrt{3}}(0, 1, 0, -1, 0, 1)\,, \nonumber\\
a' = {1\over\sqrt{3}}(c_\omega, -s_\omega, c_\omega, -s_\omega, 1, 0)\,, &&
b' = {1\over\sqrt{3}}(s_\omega, c_\omega, -s_\omega, -c_\omega, 0, 1)\,, \nonumber\\
a'' = {1\over\sqrt{3}}(c_\omega, s_\omega, c_\omega, s_\omega, 1, 0)\,, &&
b'' = {1\over\sqrt{3}}(-s_\omega, c_\omega, s_\omega, -c_\omega, 0, 1)\,,
\eea
where $c_\omega \equiv \Re\,\omega = -1/2$, $s_\omega \equiv \Im\,\omega = \sqrt{3}/2$, 
and each $(a, b)$-pair corresponds to the same eigenvalue.
Each pair of these vectors satisfies $D^{(aa)} = - D^{(bb)} = a$,
which coincides with Check-(3)(8) which we used above for detection 
of the $e^{(8)}$ and $e^{(3)}$ eigenvectors.
Therefore, we arrive at remarkable simple basis-invariant condition for the 
$CP$-violating $\Delta(54)$-symmetric 3HDM:
\begin{itemize}
\item
the eigenvalues of the matrix $\Lambda$ display the degeneracy pattern $2+2+2+2$;
\item
for each eigenvalue, the two eigenvectors $a, b$ pass Check-(3)(8);
\item
vectors $M=0$, $L=0$.
\end{itemize}
The $CP$-conserving case corresponds to the situation 
where the four pairs of eigenvectors exhibit the above properties 
but the eigenvalue degeneracy pattern becomes $2+2+4$.

Finally, the largest discrete symmetry group which can be imposed on the 3HDM scalar sector
is $\Sigma(36)$, which is twice larger than $\Delta(54)$.\footnote{The notation $\Sigma(36)$
indicates the subgroup of $PSU(3)$, which becomes $\Sigma(36\varphi)$, the group of order 108, within $SU(3)$.}
It arises in the real $\bar\lambda$ basis
if the coefficients of $V_{\Delta(54)}$ satisfy an additional constraint: $3\lambda_{38} = \lambda' + 2 \bar\lambda$.
The potential then becomes symmetric under the following transformation of order 4:
\be
d = {1 \over \sqrt{3}}\mmmatrix{1}{1}{1}{1}{\omega^2}{\omega}{1}{\omega}{\omega^2}\,,
\ee
such that $d^2$ describes the transposition of $\phi_2 \leftrightarrow\phi_3$. Adding $d$ to the
symmetry generators leads to $\Sigma(36) \simeq (\Z_3 \times \Z_3)\rtimes \Z_4$.
The basis-invariant path to this symmetry group is to observe the four pairs of eigenvectors 
satisfying the same conditions as for the $\Delta(54)$-case, but with the eigenvalue degeneracy pattern $4+4$.


\section{Conclusions and outlook}

In this paper, we solved the notoriously difficult problem of recognizing in a basis-independent way 
whether a 3HDM scalar potential has a symmetry. Similar methods for 2HDM existed for more than a decade,
but generalizing them beyond two doublets proved challenging.
Within 3HDM, prior to this work, it was known which symmetry groups $G$
can be imposed on its scalar sector and how to write general potentials
invariant under each $G$ in a {\em special basis}, in which the generators of $G$ take simple form.
However it was always understood that if the same $G$-symmetric 3HDM was written
in a different basis, the presence of $G$ would be hidden and recognizing it would become very challenging.

Developing the ideas suggested very recently in \cite{Ivanov:2018ime} and \cite{Ivanov:2019kyh},
we constructed a novel formalism which efficiently detects structural properties
of 3HDM scalar sectors in any basis.
The key role is played by the constructions in the adjoint space of the $SU(3)$ basis transformation group,
and specifically by the products of the adjoint-space vectors based on the $SU(3)$-invariant tensors
$f_{ijk}$ and $d_{ijk}$.

Despite being technical, the results of this paper remove an important obstacle on the road
towards efficient phenomenological exploration of 3HDMs.
When performing a scan over the scalar parameter space,
one can now detect not only the symmetry group 
but also {\em proximity} of a model to a symmetric situation.
Since various symmetry groups can lead to certain patterns in the scalar and flavour sectors,
all models sufficiently close to these symmetric cases will inherit some of these features.
This proximity can now be detected irrespective of basis choice.

This is particularly important for models which contain not only three Higgs doublets 
equipped with a symmetry group $G$ but also additional fields.
The loop corrections by these fields can modify effective Higgs self-couplings,
shifting the model in the parameter space away from the chosen $G$-symmetric point.
With the results of this work, one can quantify this shift in basis-independent way.

One can also investigate situations when a model is close to several symmetry situations simultaneously.
In this case, one may observe and explore competing effects of proximity to the two symmetry groups.
Such studies will generate not only numerical results but also a qualitative intuition
of how one should build multi-Higgs-doublet models with desired phenomenological properties.

\section*{Acknowledgements}
We thank Celso~Nishi, Jo\~{a}o~P.\ Silva, and Andreas Trautner for many useful discussions and numerous comments on the paper.

We acknowledge funding from the Portuguese
\textit{Fun\-da\-\c{c}\~{a}o para a Ci\^{e}ncia e a Tecnologia} (FCT) through the FCT Investigator 
contracts IF/00989/2014/CP1214/CT0004 and IF/00816/2015
and through the projects UID/FIS/00777/2013, CERN/FIS-NUC/0010/2015, and PTDC/FIS-PAR/29436/2017,
which are partially funded through POCTI (FEDER), COMPETE, QREN, and the EU.
We also acknowledge the support from National Science Center, Poland, via the project Harmonia (UMO-2015/18/M/ST2/00518).

\appendix

\section{Detecting subspaces}\label{appendix-subspaces}

We showed in section~\ref{detecting-subspaces} that the products of adjoint space vectors $a$ and $b$
\be
F^{(ab)}_i \equiv f_{ijk} a_j b_k\,,\quad D^{(ab)}_i \equiv \sqrt{3} d_{ijk} a_j b_k\,,
\quad D^{(aa)}_i \equiv \sqrt{3} d_{ijk} a_j a_k\,.
\label{fd-products-2}
\ee
can be used to identify basis-invariant features of the subspaces to which these vectors belong.
These products satisfy relations \eqref{DD} and \eqref{DF1}, which we now rewrite assuming 
vectors $a$ and $b$ are orthonormal:
\be
D^{(aa)} D^{(ab)} = 0\,,\quad |D^{(aa)}|^2 = 1\,,\quad
D^{(aa)} D^{(bb)} + 2 |D^{(ab)}|^2 = 1\,,\quad |D^{(ab)}|^2 + |F^{(ab)}|^2 = 1\,.
\label{DDD}
\ee
When applied to the eigenvectors of $\Lambda$, this technique can ensure
that in an appropriate basis $\Lambda$ has a block-diagonal form.

\subsection{1D and 2D subspaces}\label{detecting-1D2D}

The two examples given in the main text correspond to basis-invariant detection
of 1D and 2D subspaces. Let us summarize them here for completeness.
\begin{itemize}
\item
If a vector $a$ satisfies $D^{(aa)} = -a$, then there exists a basis
in which $a$ is aligned with $+x_8$ direction.
When applied to the eigenvectors of $\Lambda$, this requirement constitutes Check-(8).
No other basis-invariant condition detecting an 1D subspace with different properties exists.
\item
If two vectors $a$ and $b$ satisfy $F^{(ab)} = 0$, or alternatively $D^{(aa)} = -D^{(bb)}$,
then there exists a basis in which they both lie in the $(x_3,x_8)$ subspace. 
If, in addition, one observes that $D^{(aa)} = -D^{(bb)} = -a$,
then these vectors are aligned with $x_8$ and $x_3$, respectively.
When applied to the eigenvectors of $\Lambda$, these two versions of the conditions 
give Check-(38) and Check-(3)(8), respectively.
Notice also that if it happens that two eigenvectors passing Check-(38) 
correspond to the same eigenvalue, one can always find their linear combinations
which will pass Check-(3)(8).
\end{itemize}

\subsection{Detecting 3D subspaces}\label{detecting-3D}

As we already described in section~\ref{detecting-subspaces},
having identified an eigenvector $e^{(8)}$ via Check-(8),
we can easily detect if there are three other eigenvectors spanning 
the subspace $(x_1,x_2,x_3)$; this was formulated as Check-(123)(8).
However, it is also possible to detect three eigenvectors from this 3D subspace
even without the presence of $e^{(8)}$.

Suppose one has three orthonormal vectors $a, b, c$ which are closed under $f$-product:
\be
\boxed{
F^{(ab)} = c\,, \quad F^{(bc)} = a\,, \quad F^{(ca)} = b\,.\label{space123}
}
\ee
Then their respective hermitian matrices $A$, $B$, $C$ form the $su(2)$ subalgebra of $su(3)$.
It implies that, back in the adjoint space, one can always rotate them to the space $(x_1, x_2, x_3)$
and, if needed, align them with the axes.
This observation is the basis of {\bf Check-(123)}: if one finds three mutually orthogonal eigenvectors
of $\Lambda$ which obey \eqref{space123}, then there exist a basis 
in which $\Lambda$ has a $3\times 3$ block in $(x_1,x_2,x_3)$,
and, moreover, this block can be diagonalized.

Using the relations \eqref{DDD}, one can reformulate the conditions \eqref{space123}
in terms of $D$'s. Indeed, since the three vectors $F$'s have unit absolute values, 
their respective $D^{(ab)}=D^{(bc)}=D^{(ac)}=0$, and as a result we observe
\be
D^{(aa)}=D^{(bb)}=D^{(cc)}\,.\label{space123-D}
\ee
Notice that this version of Check-(123) may be easier to verify than \eqref{space123} because
one is not forced to test all pairs of eigenvectors.

The conserve is also true: if three orthonormal vectors $a, b, c$ satisfy \eqref{space123-D},
then one can rotate them to the $(x_1,x_2,x_3)$ subspace.
Indeed, from the relations \eqref{DDD} one concludes that $D^{(ab)}=D^{(bc)}=D^{(ac)}=0$.
This means that the three corresponding hermitian traceless matrices $A$, $B$, $C$
anticommute with each other. Thus, they form the 3D Clifford algebra and,
despite being $3\times 3$ matrices, they can be expressed as Pauli matrices 
within a $2\times 2$ block and zeros otherwise. Back in the adjoint space, 
this means that $a, b, c$ are located in the $(x_1,x_2,x_3)$ subspace.

It is also possible that the three orthonormal vectors $a, b, c$, which
are closed under $f$-product, need to be corrected by the factor 2: 
\be
\boxed{
2F^{(ab)} = c\,, \quad 2F^{(bc)} = a\,, \quad 2F^{(ca)} = b\,.\label{space257}
}
\ee
Then, the matrices $A$, $B$, $C$ form the $so(3)$ subalgebra of $su(3)$. 
One can always rotate the three vectors to the subspace $(x_2, x_5, x_7)$
or to other equivalent subspaces such as $(x_2, x_4, x_6)$, etc. 
This property is the basis of {\bf Check-(257)}, which 
was used in \cite{Nishi:2006tg} to detects explicit $CP$ conservation in 3HDM.

\subsection{Detecting 4D subspaces}\label{detecting-4D}

A direct inspection of the non-zero elements of the tensors $f_{ijk}$ and $d_{ijk}$ 
given in Eqs.~\eqref{tensor-fijk} and \eqref{tensor-dijk} reveals that
they contain an odd number of indices from the set $(1,2,3,8)$ 
and an even number of indices from the set $(4,5,6,7)$.\footnote{%
The similar observation applies to the splitting $(3,6,7,8)$ vs. $(1,2,4,5)$, 
and to the splitting $(3,4,5,8)$ vs. $(1,2,6,7)$, which differ just by Higgs doublet permutation.
For definiteness, we focus on the first splitting.}
Taking any four orthonormal vectors $a, b, c, d \in (x_1, x_2, x_3, x_8)$, we observe
that their $f$-products stay within the same subspace. 
Therefore, the corresponding hermitian matrices $A$, $B$, $C$, $D$
form a 4D subalgebra of $su(2)\times u(1) \subset su(3)$.

Conversely, if we observe that four orthonormal vectors $a, b, c, d$ are such that
all their $f$-products lie in the same 4D space spanned by $a, b, c, d$, then their hermitian matrices 
form a 4D subalgebra of $su(3)$, which can only be $su(2)\times u(1)$.
Therefore, there exists a basis, in which vectors $a, b, c, d$ lie in $(x_1, x_2, x_3, x_8)$.

Applying this observation to the eigenvectors of $\Lambda$, 
we obtain {\bf Check-(1238)}, or equivalently {\bf Check-(4567)}:
if $\Lambda$ has four mutually orthogonal eigenvectors whose $f$-products
lie in the same 4D subspace, then and only then
there exists a basis in which $\Lambda$ takes the block-diagonal form
with two $4\times 4$ blocks, one lying in $(x_1, x_2, x_3, x_8)$
and the other lying in $(x_4, x_5, x_6, x_7)$.

\section{Splitting $V_6$}\label{appendix-B}

If $\Lambda$ passes Check-(38), it takes, in an appropriate basis, a block-diagonal
form with a $2\times 2$ block within $(x_3,x_8)$ and
a $6\times 6$ block within the subspace $V_6 = (x_1,x_2,x_4,x_5,x_6,x_7)$.
In certain symmetry constrained cases, this block can be split further
or can exhibit specific patterns.
Here, we investigate the relevant options and give their basis-invariant conditions.

\subsection{Detecting $2\times 2$ blocks}\label{splitting-V6-D}

Since Check-(38) is passed, we already have a pair of eigenvectors
which define the $(x_3,x_8)$ subspace.
Let us now pick up two orthonormal vectors $a, b \in V_6$. If their products satisfy
\be
\boxed{
D^{(ab)} = 0\quad \mbox{and} \quad D^{(aa)} = D^{(bb)} \in (x_3,x_8)\,,
}
\label{Check-12}
\ee
then, as we prove below, there exists a basis, in which the vectors $a$ and $b$ 
lie within subspace $(x_1,x_2)$ or $(x_4,x_5)$ or $(x_6,x_7)$.
This feature is the basis of {\bf Check-(12)}:
if, after passing Check-(38), the matrix $\Lambda$ has two eigenvectors within $V_6$ 
which satisfy \eqref{Check-12}, then its has a $2\times 2$ block located within subspace $(x_1,x_2)$
or $(x_4,x_5)$ or $(x_6,x_7)$.

The proof goes as follows.
Denote $a = (a_1,a_2, a_4,a_5,a_6,a_7)$ and $b = (b_1,b_2, b_4,b_5,b_6,b_7)$
and compute the $D$-products explicitly.
First, write down $D^{(ab)}_3 = 0$, $D^{(ab)}_8 = 0$:
\bea
&&(a_4b_4+a_5b_5) - (a_6b_6+a_7b_7) = 0\,, \nonumber\\
&&2 (a_1b_1+a_2b_2)  - (a_4b_4+a_5b_5)  - (a_6b_6+a_7b_7)  = 0\,.
\eea
Together with the orthogonality condition $\vec a\vec b = 0$, they lead to
\be
a_1b_1+a_2b_2 = a_4b_4+a_5b_5 = a_6b_6+a_7b_7 =0\,.
\ee
This implies the following structure for $b$:
\be
b = (-\sigma a_2, \sigma a_1, \sigma'\, a_5, -\sigma'\,a_4, -\sigma'' a_7, \sigma'' a_6)\,,\label{vector-b-step1}
\ee
with some real coefficients $\sigma$, $\sigma'$, $\sigma''$.

Next, from $D^{(aa)} = D^{(bb)}$ within the subspace $(x_3,x_8)$ as well as from the normalization condition
$\vec a^2 = \vec b^2 = 1$, we see that $\sigma$'s can only be $\pm 1$.

Finally, using this form of $b$, let us explicitly write $D^{(ab)}$, $D^{(aa)}$ and $D^{(bb)}$ within $V_6$:
\be
D^{(ab)} = -{\sqrt{3} \over 2}\left(\begin{array}{c}
(\sigma'+\sigma'')(a_4a_7-a_5a_6) \\
(\sigma'+\sigma'')(a_4a_6 + a_5a_7) \\
(\sigma+\sigma'')(a_1a_7 + a_2a_6) \\
(\sigma+\sigma'')(-a_1a_6 + a_2a_7) \\
(\sigma+\sigma')(-a_1a_5 + a_2a_4) \\
(\sigma+\sigma')(a_1a_4+a_2a_5)
\end{array}\right)\,,\label{Dab-V6}
\ee
and 
\be
D^{(aa)} = \sqrt{3} \left(\begin{array}{c}
a_4a_6+a_5a_7 \\
-a_4a_7 + a_5a_6 \\
a_1a_6 - a_2a_7 \\
a_1a_7 + a_2a_6\\
a_1a_4 + a_2a_5 \\
a_1a_5 - a_2a_4
\end{array}\right)\,,\quad
D^{(bb)} = - \sqrt{3} \left(\begin{array}{c}
\sigma'\sigma''(a_4a_6+a_5a_7) \\
\sigma'\sigma''(-a_4a_7 + a_5a_6) \\
\sigma\sigma''(a_1a_6 - a_2a_7) \\
\sigma\sigma''(a_1a_7 + a_2a_6)\\
\sigma\sigma'(a_1a_4 + a_2a_5) \\
\sigma\sigma'(a_1a_5 - a_2a_4)
\end{array}\right)\,.\label{DaaDbb-V6}
\ee
Setting $D^{(aa)} = 0$ within $V_6$ implies that among the three pairs $(a_1,a_2)$, $(a_4,a_5)$, and $(a_6,a_7)$
only one can be non-zero, and the same applies to $b$. Thus, vectors $a$ and $b$ 
are located within one of these three blocks.
If they are eigenvectors of $\Lambda$, it implies that the corresponding
$2\times 2$ block is decoupled from the rest. Notice that by permuting the Higgs doublets,
one can always make this block to lie within the $(x_1,x_2)$ subspace.

If {\em two pairs} of eigenvectors from $V_6$ pass Check-(12),
then the entire $6\times 6$ matrix $\Lambda$ within $V_6$ is split in three 
$2\times 2$ blocks located within subspace $(x_1,x_2)$
or $(x_4,x_5)$ or $(x_6,x_7)$.
If $\Lambda$ has this property, we say it passes {\bf Check-(12)(45)(67)}.

\subsection{$Z_3$ pattern inside $V_6$}\label{Z3-pattern}

Let us relax the conditions \eqref{Check-12} which defined Check-(12) and
require now that
\be
\boxed{
D^{(ab)} \in V_6\,, \quad D^{(aa)} - D^{(bb)} \in V_6\,,
\quad D^{(aa)} + D^{(bb)} \in (x_3,x_8)\,,
}
\label{Check-Z3}
\ee
which we call {\bf Check-$\Z_3$}.
That is, we now allow for non-zero vectors $D^{(ab)}$ and $D^{(aa)} - D^{(bb)}$
provided they belong to $V_6$.
Repeating the calculations of section~\ref{splitting-V6-D},
we see that all components of $a$ and $b$ can be non-zero.
However $b$ must still be of the form \eqref{vector-b-step1}
with $\sigma=\sigma'=\sigma'' = \pm 1$.

Next, suppose the two eigenvectors of $\Lambda$, which we denote $e$ and $e'$, 
satisfy \eqref{Check-Z3} and correspond to the same eigenvalue $\lambda$. 
It can be immediately checked that
their contribution to the eigensystem expansion for $\Lambda$, $e_i e_j + e'_i e'_j$, has the following form: 
\be
\left(
\begin{array}{cccccc}
 c_{12} & 0 & g_6 & h_6 & g_7 & h_7 \\
 0 & c_{12} & h_6 & -g_6 & -h_7 & g_7 \\
 g_6 & h_6 & c_{13} & 0 & g_5 & -h_5 \\
h_6 & -g_6 & 0 & c_{13} & -h_5 & -g_5 \\
g_7 & -h_7 & g_5 & -h_5 & c_{23} & 0 \\
h_7 & g_7 & -h_5 & -g_5 & 0 & c_{23} \\
\end{array}\label{Z3-block6x6-2}
\right)
\ee
It is remarkable that this block has exactly the same form as in the $\Z_3$-symmetric 3HDM, Eq.~\eqref{Z3-block6x6}.
Therefore, if the eigenvalues of $\Lambda$ within $V_6$ are pairwise degenerate,
and if the three corresponding pairs of eigenvectors satisfy Check-$\Z_3$ given in Eq.~\eqref{Check-Z3},
then we obtain the $\Z_3$-symmetric model.
Notice that the three pairs of eigenvectors may be in arbitrary orientation with respect to each other;
apart from mutual orthogonality, there are no constraints. 

To prove the converse statement, we notice that the $6\times 6$ block \eqref{Z3-block6x6-2}
keeps its structural form when raised to any power.
It has three pairwise degenerate eigenvalues, therefore it can be written
generically as 
\be
\lambda_1(e_{1i}e_{1j} + e'_{1i}e'_{1j}) + 
\lambda_2(e_{2i}e_{2j} + e'_{2i}e'_{2j}) + \lambda_3(e_{3i}e_{3j} + e'_{3i}e'_{3j})\,.
\ee
Its square and cube have the same form with squared and cubed eigenvalues, respectively.
This can happen only if each eigensystem $e_i e_j + e'_i e'_j$ has the form \eqref{Z3-block6x6-2}. 
Contracting it with $\sqrt{3}d_{ijk}$ gives the vector $D^{(ee)}+D^{(e'e')}$, 
and one can verify by explicit calculation that it indeed belongs to $(x_3,x_8)$.

Next, we checked with \verb|Mathematica| that each pair of eigenvectors $(e,e')$ of this matrix
has the form of vectors $a$ and $b$ as in \eqref{vector-b-step1}.
That is, not only are the eigenvectors $(e,e')$ themselves orthogonal and equally normalized
but so are their 2D components within the subspaces $(x_1,x_2)$, $(x_4,x_5)$ and $(x_6,x_7)$.
This immediately implies that $D^{(ab)}$ and $D^{(ee)}-D^{(e'e')}$ 
cannot have any components in the $(x_3,x_8)$.
Thus, we arrive at all three conditions of Check-$\Z_3$ in \eqref{Check-Z3}.

\subsection{$U(1)$ patterns inside $V_4$}\label{structures-V4}

Suppose two vectors $a, b \in V_4 = (x_4,x_5,x_6,x_7)$. 
By inspecting entries of the tensor $d_{ijk}$, 
one sees that $D^{(ab)}$, $D^{(aa)}$, and $D^{(bb)}$ 
must all lie in the subspace $(x_1,x_2,x_3,x_8)$.
In this situation, let us now impose a requirement similar to \eqref{Check-12}:
\be
\boxed{
D^{(ab)} = 0\quad \mbox{and} \quad D^{(aa)} = D^{(bb)}\,.
}
\label{Check-U12}
\ee
Then, one can establish by direct computation that for any 
$a = (a_4,a_5,a_6,a_7)$ one can pick up the vector $b = (a_5, - a_4, a_7, -a_6)$ to satisfy \eqref{Check-U12}. 

Now, suppose $\Lambda$ has passed Check-(4567)
and, within the subspace $V_4$, it has two pairs of eigenvectors 
which satisfy \eqref{Check-U12}. Then we say is passed {\bf Check-$U(1)_2$}.
Writing $\Lambda$ via the eigensystem expansion, we get the $4\times 4$ block of the following form:
\be
\mtrx{
c_{13} & 0 & g_8 & h_8 \\
0 & c_{13} & -h_8 & g_8 \\
g_8 & -h_8 & c_{23} & 0 \\
h_8 & g_8 & 0 & c_{23}}\,.\label{U12-block4x4-2}
\ee
This form is exactly what we obtained for the 3HDM invariant under $U(1)_2$ symmetry.

Alternatively, we can also impose a different condition:
\be
\boxed{
D^{(ab)} =0\ \mbox{in}\ (x_3,x_8)\,, \quad D^{(aa)} = D^{(bb)}\ \mbox{in} \ (x_3,x_8)\,,
\quad D^{(aa)} = - D^{(bb)}\ \mbox{in} \ (x_1,x_2)\,.
}
\label{Check-U11}
\ee
Notice that, unlike the previously considered example, this set of conditions
explicitly distinguishes subspaces $(x_1,x_2)$ and $(x_3,x_8)$.
Thus, it can be used only after we have already passed Check-(38) and Check-(12).

Now, once again, suppose that within $V_4$, matrix $\Lambda$ has two degenerate eigenvalues
each corresponding to a pair of eigenvectors which satisfy \eqref{Check-U11}.
Then we say $\Lambda$ passes {\bf Check-$U(1)_1$}.
The $4\times 4$ block constructed via the eigensystem expansion now has the following form:
\be
\mtrx{
c_{13} & 0 & g_5 & h_5 \\
0 & c_{13} & h_5 & -g_5 \\
g_5 & h_5 & c_{23} & 0 \\
h_5 & -g_5 & 0 & c_{23}}\,.\label{U11-block4x4-2}
\ee
It reproduces the corresponding block for the 3HDM invariant under $U(1)_1$ symmetry.


\end{document}